\documentclass{article}
\pdfoutput=1
\usepackage{jcappub}

\title{An improved fitting formula for the dark matter bispectrum}
\author[a,b]{H\'ector Gil-Mar\'in,}
\author[b]{Christian Wagner,}
\author[b]{Frantzeska Fragkoudi,}
\author[c,b]{Raul Jimenez}
\author[c,b]{and Licia Verde}

\affiliation[a]{Institut de Ci\`encies de l'Espai (ICE), Facultat de Ci\`encies , Campus UAB (IEEC-CSIC), Bellaterra E-08193, Spain}
\affiliation[b]{Institut de Ci\`encies del Cosmos (ICC), Universitat de Barcelona (IEEC-UB), Mart\'i  i Franqu\'es 1, E-08028, Spain}
\affiliation[c]{ICREA Instituci\'o Catalana de Recerca i Estudis Avan\c{c}ats. Passeig Llu\'is Companys 23,
E-08010 Barcelona, Spain}

\emailAdd{gil@ieec.uab.es}
\emailAdd{cwagner@icc.ub.edu}
\emailAdd{francesca.fragkoudi@gmail.com}
\emailAdd{raul.jimenez@icc.ub.edu}
\emailAdd{liciaverde@icc.ub.edu}

\abstract{In this paper we present an improved fitting formula for the dark matter bispectrum  motivated by the previous phenomenological approach of Ref \cite{SC}. We use a set of LCDM simulations to calibrate the fitting parameters in the $k$-range of $0.03\, h/\mbox{Mpc} \leq k \leq 0.4\, h/\mbox{Mpc}$ and in the redshift range of $0\leq z \leq 1.5$. This new proposed fit describes well the BAO-features although it was not designed to. The deviation between the  simulations output  and our analytic prediction is typically less than 5\% and in the worst case is never above 10\%. We envision that this new analytic fitting formula will be very useful in providing reliable predictions for the non-linear dark matter bispectrum for LCDM models.}

\begin{document}

\maketitle


\section{Introduction}

The dark matter and galaxy power spectrum have been widely used to study the growth of structure, to constrain cosmological parameters and galaxy bias models. These tools have proved very successful  and have contributed to crystallize the current LCDM model  e.g., \cite{sdss_data} and refs therein.
With ongoing and forthcoming galaxy surveys, like BOSS\footnote{Baryon Oscillation Spectroscopic Survey} and EUCLID\footnote{R. Laurejis et al, arXiv:1110.3193}, the signal-to-noise of the data will increase and the uncertainties around this model will be reduced. Higher precision data will allow the use of not only the two-point correlation function, but also of higher-order statistics, in order to constrain and improve our theories and models. The bispectrum (the three-point correlation function in Fourier space) is naturally the next statistic to consider \cite{FryMelott85, kayo}. Using both the power spectrum and bispectrum we can improve our knowledge of the growth of structure and galaxy biasing \cite{Fry94, Fryetal95, MVH97, VHM98, guo2, Verde2df01,Scoccimarroetal01,Feldmanetal01,pollack},  constrain possible departures from Gaussianity in the initial conditions of the matter density field \cite{VWHK00, VJKM01, Scoccimarroetal04, SefusattiKomatsu07, jeong} as well as constrain departures from GR e.g., \cite{Shirataetal07,hgm}.

From a theoretical point of view, perturbation theory  and subsequent improvements such as renormalized perturbation theory \cite{CrocceScoccimarro}, resummed perturbation theory or time-RG flow \cite{Pietroni2} is  a physically well-motivated approach to study these statistical moments. Tree-level perturbation theory has demonstrated to describe well the behavior of the power spectrum and bispectrum at large scales. However, non-trivial computations are needed to obtain predictions at non-linear scales: the one-loop correction and beyond,  for the power spectrum and bispectrum. For the power spectrum, however, other phenomenological approaches have been demonstrated to work better for a wide range of redshifts and different cosmologies  e.g., \cite{halofit, halomodel,MaFry00}. For the bispectrum there are also simple phenomenological models that predict its  behavior at non-linear scales \cite{SC}, but they fail to accurately reproduce  the BAO-features \cite{PCS} and are only precise at the 20\%-30\% level. Therefore better analytical models are needed to describe the bispectrum at these non-linear scales.

In this paper we improve the phenomenological description presented by \cite{SC} (hereafter SC) more than 10 years ago. 
Using a set of modern simulations we fit the free parameters of our proposed analytic formula. Thus, we obtain an improved description for the bispectrum in the LCDM model scenario (including baryonic acoustic oscillations) in a range of $0.03\, h/\mbox{Mpc} \leq k \leq 0.4\, h/\mbox{Mpc}$ and for different redshifts, $0\leq z \leq 1.5$.

This paper is organized as follows: in \S 2 we begin with a description of the density field statistics and different analytic approaches to the dark matter bispectrum.
In \S 3 we describe the simulations we use to fit the parameters. In \S 4 we present our results, compare them with previous fitting formulae and with 1-loop corrections and discuss the differences. We finally conclude in \S 5.
In  Appendix \ref{Appendix_A}, we give details of how the bispectrum and its errors are computed from simulations. In Appendix \ref{Appendix_B} we test how our formula works for other non-standard LCDM models. In Appendix \ref{Appendix_C} we present a short description of 1-loop correction in Eulerian perturbation theory.
\section{Theory}

\subsection{Power spectrum \& bispectrum}

The power spectrum $P(k)$, the Fourier transform of the two-point correlation function, is one of the simplest statistics of interest one can extract from the dark matter overdensity field $\delta({\bf k})$,
\begin{equation}
 \langle \delta({\bf k})\delta({\bf k'})\rangle \equiv (2\pi)^3 \delta^D({\bf k+k'}) P(k)\,,
\label{Pk}
\end{equation}
where $\delta^D$ denotes the Dirac delta function and $\langle\dots \rangle$ the ensemble average over different realizations of the Universe. Under the assumption of an isotropic Universe, the power spectrum does not depend on the direction of the $\bf k$-vector. Since we only have one observable Universe, the average $\langle\dots \rangle$ is taken over all different directions  for each $\bf k$-vector. Under the hypothesis of ergodicity both averages will yield the same result.

The second statistic of interest is the bispectrum $B$, defined by,
\begin{equation}
 \langle \delta({\bf k}_1) \delta({\bf k}_2) \delta({\bf k}_3)\rangle\equiv (2\pi)^3 \delta^D({\bf k}_1+{\bf k}_2+{\bf k}_3) B({\bf k}_1, {\bf k}_2, {\bf k}_3).
\label{Bk}
\end{equation}
The Dirac delta function ensures that the bispectrum is defined only for ${\bf k}$-vector configurations that form closed triangles: $\sum_i {\bf k}_i=0$. Note that $\delta({\bf k}_1) \delta({\bf k}_2) \delta({\bf k}_3)$ is in general a complex number, however once the average is taken, the imaginary part goes to zero.

It is convenient to define the reduced bispectrum $Q_{123}\equiv Q({\bf k}_1, {\bf k}_2, \bf{k}_3)$ as,
\begin{equation}
Q_{123}\equiv \frac{B({\bf k}_1, {\bf k}_2, {\bf k}_3)}{P(k_1)P(k_2)+P(k_1)P(k_3)+P(k_2)P(k_3)}
\end{equation}
which takes away part of the dependence on scale and cosmology\footnote{For equilateral configuration and up to tree level, $Q$ does not depend on cosmology or scale.}. The reduced bispectrum is useful when comparing different models, since it only has a weak dependence on cosmology and one can thus break degeneracies between cosmological parameters in order to isolate the effects of gravity. 

The bispectrum for Gaussian initial conditions is zero and remains zero in linear theory, i.e. as long as the $k$-modes evolve independently\footnote{Wick theorem states that the $n$-point  correlation function of a Gaussian field is always zero when $n$ is an odd number.}. However, when non-linearities start to play an important role, mode coupling is no longer negligible and  the bispectrum becomes non-zero. Thus, by measuring the bispectrum one can extract information about how non-linear processes influence the evolution of dark matter clustering.

\subsection{Analytic approaches in the literature}
In order to understand the observational data, we need accurate theoretical predictions for $B(k_1,k_2,k_3)$. A physically well-motivated analytic theory for doing this, is perturbation theory (PT hereafter) (see \cite{PT_review} for a review) or subsequent improvement such as renormalized PT, resummed PT etc.

In an Einstein de-Sitter Universe (hereafter EdS Universe) and at second order (tree-level) in Eulerian perturbation theory, the bispectrum is given by \citep{Fry84},
\begin{equation}
 B_{123}=2F^s_2({\bf k}_1,{\bf k}_2) P^L_1 P^L_2 + \mbox{cyc. perm.},
\label{PT}
\end{equation}
where $B_{123}=B({\bf k}_1, {\bf k}_2, {\bf k}_3)$, $P^L_i=P^L(k_i)$ is the linear power spectrum, and the symmetrized two-point kernel $F^s_2$ is given by      
\begin{equation}
 F^s_2({\bf k}_i,{\bf k}_j)= \frac{5}{7}+\frac{1}{2}\cos(\theta_{ij})\left(\frac{k_i}{k_j}+\frac{k_j}{k_i}\right)+\frac{2}{7}\cos^2(\theta_{ij}),
 \label{kernel}
\end{equation}
where $\theta_{ij}$ is the angle between the vectors ${\bf k}_i$ and ${\bf k}_j$.
This formula is the second order perturbation theory contribution to the bispectrum which is the leading order contribution. On quasi-linear scales, this expression is a very good prediction but fails in the moderate non-linear regime. The dependence on cosmology of the two-point kernel $F^s_2$ is very weak and hence the cosmology dependence of the bispectrum is almost completely contained in $P^L_i$. Because of this, in this work, we use the kernel of Eq. \ref{kernel} even though we are dealing with the  LCDM model.

One can improve the tree-level PT prediction by going one step further and including one-loop corrections. However, at this point the computation of the bispectrum becomes cumbersome. For an initially Gaussian $\delta$-field this yields four additional terms to the tree-level contribution (see Appendix \ref{Appendix_C} for details).

An alternative way of reaching these non-linear scales, without using the one-loop correction, and to even push beyond the one-loop regime of validity, is with phenomenologically motivated models. Phenomenological formulae can give simpler expressions in the non-linear regime and accurate predictions for the bispectrum. However, their physical  motivation is limited and they usually have free parameters that need to be calibrated using N-body simulations.

SC proposed a fitting formula based on the structure of the formula of Eq. \ref{PT}. It consists in replacing the linear power spectrum by the non-linear one in Eq. \ref{PT} and the EdS two-point symmetrized kernel by

\begin{eqnarray}
\label{SC_kernel}F_2^{\rm eff}({\bf k}_i,{\bf k}_j)&=&\frac{5}{7}a(n_i,k_i)a(n_j,k_j) \\ 
&+&\frac{1}{2}\cos(\theta_{ij})  \left(\frac{k_i}{k_j}+\frac{k_j}{k_i}\right)b(n_i,k_i)b(n_j,k_j) +\frac{2}{7} \cos^2(\theta_{ij})c(n_i,k_i)c(n_j,k_j),\nonumber
\end{eqnarray}
where the functions $a(n,k)$, $b(n,k)$ and $c(n,k)$ are chosen to interpolate between the tree-level results and  the hyper-extended perturbation theory regime (HEPT) \cite{hept},
\begin{eqnarray}
 \nonumber \label{abc} a(n,k)&=&\frac{1+\sigma_8^{a_6}(z)[0.7Q_3(n)]^{1/2}(q a_1)^{n+a_2}}{1+(q a_1)^{n+a_2}}, \\
 b(n,k)&=&\frac{1+0.2a_3(n+3)q^{n+3}}{1+q^{n+3.5}}, \\
\nonumber c(n,k)&=&\frac{1+4.5a_4/[1.5+(n+3)^4](q a_5)^{n+3}}{1+(q a_5)^{n+3.5}}.
\end{eqnarray}
Here $n$ is the slope of the linear power spectrum at $k$,
\begin{equation}
 \label{n}n\equiv\frac{d\log P^L(k)}{d\log k}
\end{equation}
and $q\equiv k/k_{\rm nl}$, where $k_{\rm nl}$ is the scale where non-linearities start to be important and is defined as, 
\begin{equation}
 \label{knl}\frac{k_{\rm nl}^3P^L(k_{\rm nl})}{2\pi^2}\equiv1;
\end{equation}
$a_i$ are free parameters that must be fitted using data from simulations. In particular, SC propose the values, 
\begin{equation}
\nonumber a_1=0.25,\, a_2=3.5,\, a_3=2,\,  a_4=1,\, a_5=2,\, a_6=-0.2 \,.
\end{equation}
The function $Q_3(n)$ is given by
\begin{equation}
 Q_3(n)=\frac{4-2^n}{1+2^{n+1}}\,.
\label{eq:Q3}
\end{equation}
With all these changes, the SC approach reads, 
\begin{equation}
 B_{123}=2F^{\mbox{eff}}_2({\bf k}_1,{\bf k}_2) P_1 P_2 + \mbox{cyc. perm.},
\label{SC_formula}
\end{equation}
where $P_i$ is the non-linear power spectrum at $k_i$. On large scales, where the functions $a$, $b$ and $c$ $\rightarrow1$ we recover the tree-level PT formula for the bispectrum. On the other hand, on small scales $a^2\rightarrow(7/10)Q_3$ and $b$ and $c$ $\rightarrow0$ and we obtain $Q_{123}\rightarrow Q_3(n)$, which is the prediction of HEPT.

Another approach based on phenomenological formulae, is the one presented by \cite{PCS}. The main idea is to rescale the linear formula of the bispectrum, by using some scale transformation in $k$. This way, the tree-level formula can easily be extended up to non-linear scales using the ansatz $\tilde{k_i}=\left[1+\Delta_{NL}^2(k_i)\right]^{-1/3} k_i$; where $\Delta_{NL}^2(k)=P(k)k^3/(2\pi^2)$. 


This approach has by definition the drawback that it does not preserve the BAO-features of the bispectrum. In particular,  the rescaling of $k$ produces a spurious rescaling of the peaks and troughs of the BAO wiggles that do not match with the data, producing higher deviations than the SC approach. Because of that, we do not consider this approach in this paper.

\subsection{Our analytic formula}
Our approach in this paper is inspired by the SC approach. It consists of not only refitting the $a_i$ parameters from Eq. \ref{abc} but of also modifying their expression to make it more suitable for current precision N-body data and consider the redshift range of $0\leq z \leq 1.5$. In order to do that, we use simulations with more particles, larger box sizes, and more realizations (and thus higher precision, better statistics and better error-control) with respect to previous works; we also consider snapshots at different redshifts. In order to improve the fitting precision, we also add 3 more parameters to the original model. The modified functions $\tilde{a}(n,k)$, $\tilde{b}(n,k)$, $\tilde{c}(n,k)$ then read,
\begin{eqnarray}
 \nonumber \label{abc_new} \tilde{a}(n,k)&=&\frac{1+\sigma_8^{a_6}(z)[0.7Q_3(n)]^{1/2}(q a_1)^{n+a_2}}{1+(q a_1)^{n+a_2}}, \\
 \tilde{b}(n,k)&=&\frac{1+0.2a_3(n+3)(q a_7)^{n+3+a_8}}{1+(q a_7)^{n+3.5+a_8}}, \\
\nonumber \tilde{c}(n,k)&=&\frac{1+4.5a_4/[1.5+(n+3)^4](q a_5)^{n+3+a_9}}{1+(q a_5)^{n+3.5+a_9}}.
\end{eqnarray}
Note that one recovers the original SC formulae in the limit of $a_7\rightarrow1$ and $a_8,\,a_9\rightarrow0$.

\begin{figure}
\centering
\includegraphics[clip=false, trim= 110mm 35mm 20mm 35mm,scale=0.35]{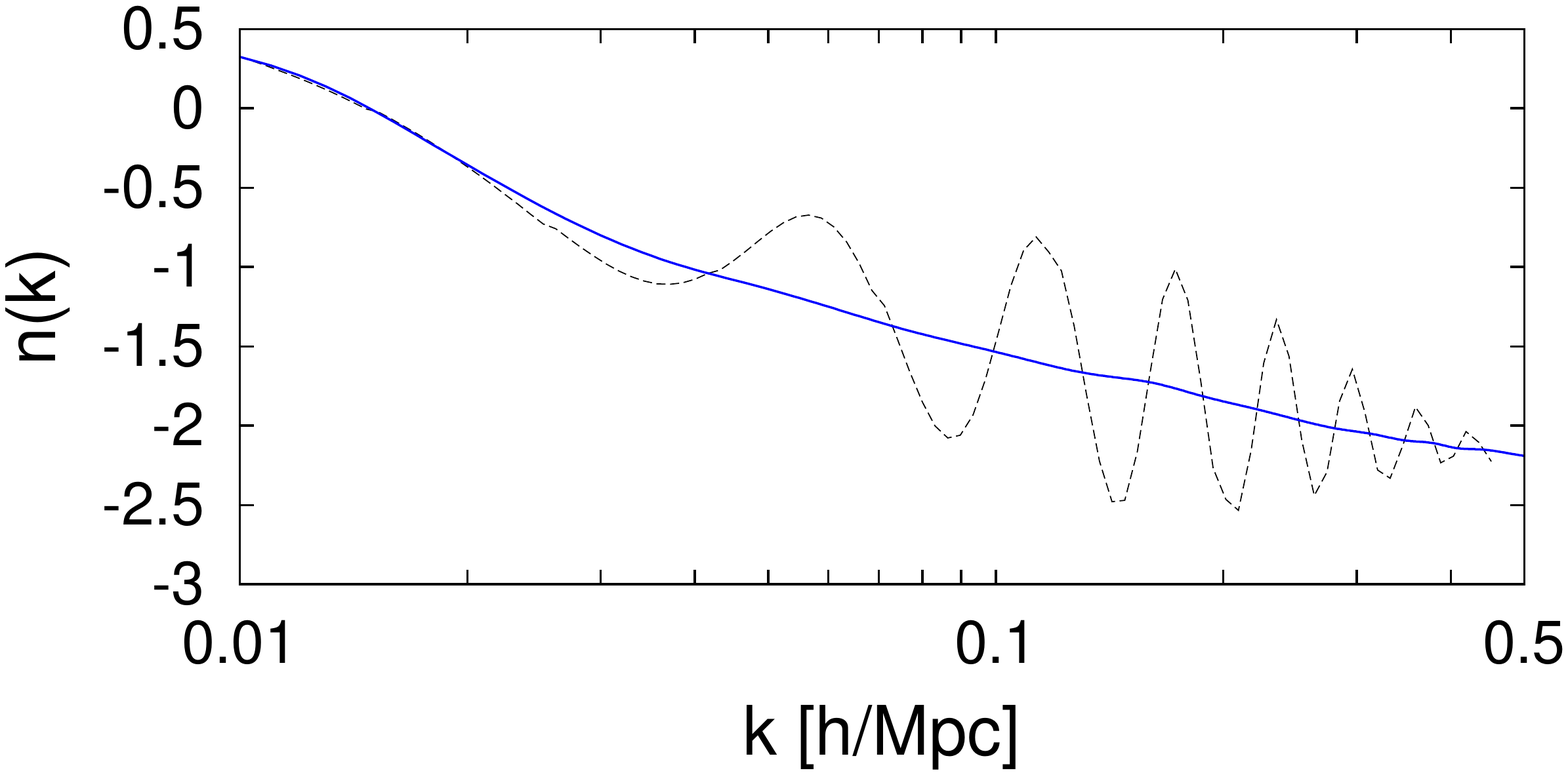}
\includegraphics[clip=false,trim= 20mm 35mm 110mm 35mm, scale=0.35]{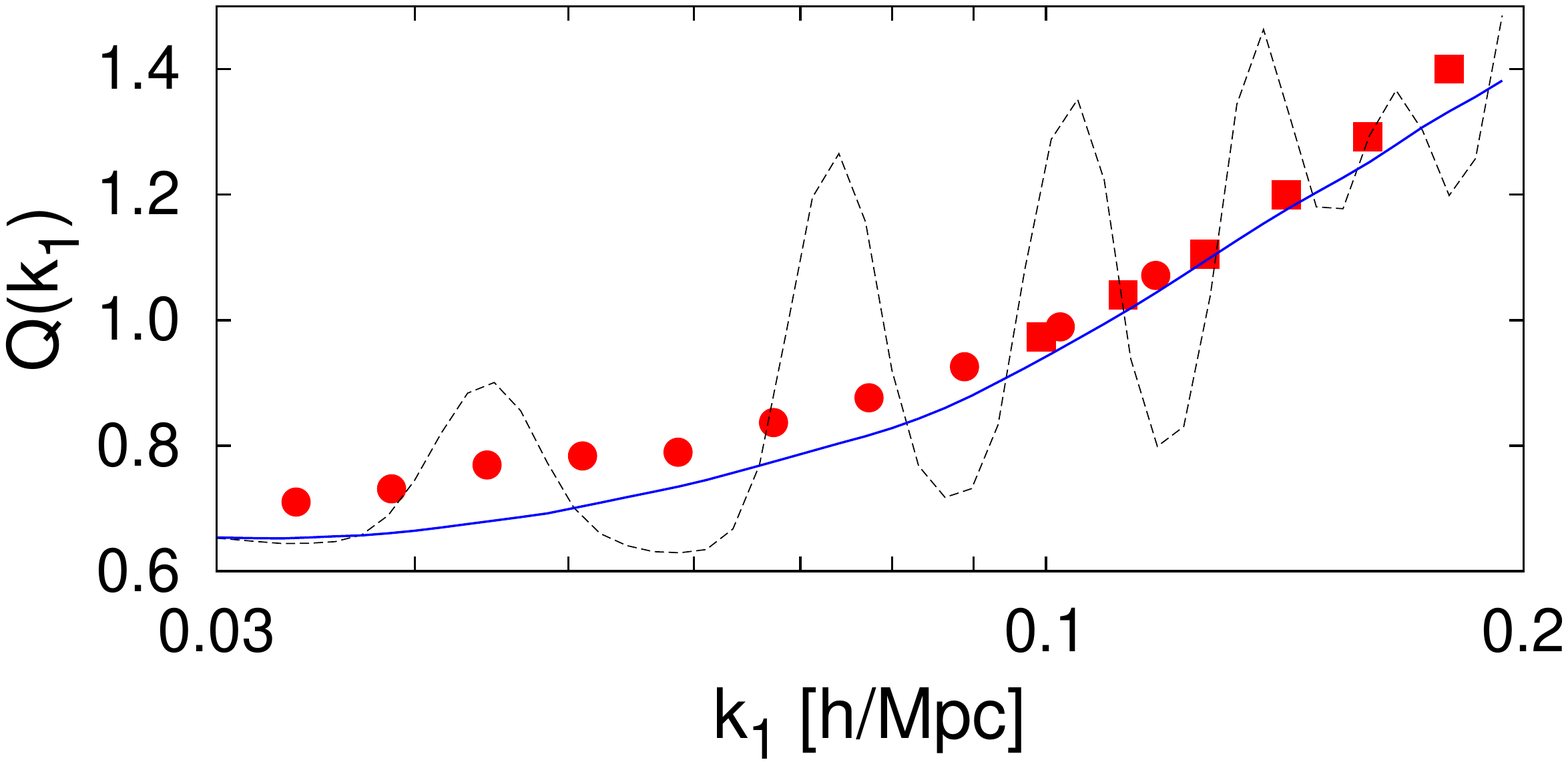}

\caption{{\it Left panel}: The slope $n(k)$ (Eq. \ref{n}) from the linear power spectrum without smoothing (black dashed line) and with a spline smoothing (blue solid line).  {\it Right panel}: $Q(k_1)$ for $k_2/k_1=2$ and $\theta_{12}=0.6\pi$. Red circles are data from simulations A and red squares from simulations B (see Table \ref{table_sims} for details on the simulations). Black dashed line is SC prediction without any spline in $n(k)$ and blue solid line with the spline in $n(k)$.} 
\label{spline_fig}
\end{figure}

The original SC formula was not designed to reproduce the BAO features. Applying this formula to a power spectrum with BAOs produces unphysical oscillations. These oscillations are much larger than those observed in simulations (see black dashed line in the right panel of Fig. \ref{spline_fig}). These oscillations are caused by the oscillatory behavior of the slope parameter $n$. 
One solution to this problem is to ``dewiggle'' the linear power spectrum \citep{sefusatti10}. However here we  want to preserve the BAO oscillations. We propose to smooth the oscillatory behavior of the parameter $n$ by means of splines, as is shown in the blue solid line of the left panel of Fig. \ref{spline_fig}. This provides an improved fit to the  BAO-features, as it is shown by the blue solid line in the right panel of Fig. \ref{spline_fig}.
In order to smooth out $n$ we calculate its spline by taking a number of points $n(k)$, 
where the points are chosen to be in the middle of the amplitude of each wiggle, such that when the 
points are connected a smooth line would pass through them. These points are used in the
spline routine, and their second order derivatives are calculated for each point $k$. This output is 
then fed into the spline routine, which returns a smoothed value of $n$ for each value of $k$.


Our method consists in using this smoothed $n$ and refit all the free $a_i$ parameters from Eq. \ref{abc_new} using the reduced bispectrum data from N-body simulations. In particular, we use the following triangle configurations at different redshifts: $\theta_{12}/\pi=0.1, 0.2,\dots,0.9$, $k_2/k_1=1.0, 1.5, 2.0, 2.5$ and $z=0, 0.5, 1, 1.5$.

\section{Simulations}\label{sims_section}

The simulations in this paper consist of two different sets, namely A and B. Each simulation is characterized by the box size, $L_b$, the number of particles, $N_p$, and the number of independent runs, $N_r$. Details about the two simulations are given in Table \ref{table_sims}.
\begin{table}
\begin{center}
\begin{tabular}{c|c|c}
& A & B \\
\hline
$L_b$ [Mpc/$h$] & 2400 & 1875 \\
\hline
$N_p$ & $768^3$ & $1024^3$ \\
\hline
$N_r$ & 40 & 3\\
\hline
$k_{N}/4$ [$h$/Mpc] & 0.25& 0.43 \\
\hline
softening $\epsilon$ [kpc/$h$] & 90 & 40 \\
\hline
PM grid  & $2048^3$ & $2048^3$ \\
\hline
ErrTolForceAcc $\alpha$  & 0.005 & 0.005  \\
\hline
initial scale factor $a_i$ & 0.05 & 0.02  \\
\hline
maximum $\Delta\log a$ & 0.025 &  0.025 \\
\hline
ErrTolIntAccuracy  $\eta$  &   0.025 &  0.025 \\
\hline
\# time steps & $\sim 1300$ & $\sim 2500$
\end{tabular}
\end{center}
\caption{Simulations details for simulations A and B. $L_b$ is the box size, $N_p$ is the number of particles, $N_r$ is the number of independent realizations. 
A quarter of the Nyquist frequency, $k_N/4$, is the upper threshold to which we trust the simulation results at the percent level. The force resolution is specified by the softening parameter $\epsilon$ and the Particle Mesh (PM) grid size. The short-range force accuracy is determined by $\alpha$ through the cell-opening criterion $M l^2 > \alpha |\mathbf{a}_{\rm old}| r^4$, where $M$ is the mass inside the cell, $l$ its side length, $\mathbf{a}_{\rm old}$ the total acceleration of the particle in the previous time step, and $r$ the distance between the particle and the cell. The remaining parameters set the time stepping in Gadget-2: the maximum global time step in the logarithm of the scale factor, $\max(\Delta\log a)$, and the parameter $\eta$ in the individual time step criterion $\Delta a = \sqrt{2\eta\epsilon/|\mathbf{a}|}$, where $\mathbf{a}$ is the acceleration of the individual particle.}
\label{table_sims}
\end{table}
 As a rule of thumb, a maximum threshold in $k$ for trusting the simulation data is set by a quarter of the Nyquist frequency, defined as $k_N/4=\pi N_p^{1/3}/(4L_b)$. At this scale it has been observed that the power spectrum starts to deviate at the 1\%-level with respect to higher resolution simulations \citep{heitmann}. We confirmed this result using our two sets of simulations.
For all the plots and results shown in this paper this limit in $k$ is never exceeded.

Both A and B simulations consist in a flat LCDM cosmology with cosmological parameters consistent with observational data. The cosmology used is $\Omega_\Lambda=0.73$, $\Omega_m=0.27$ $h=0.7$, $\Omega_bh^2=0.023$, $n_s=0.95$ and $\sigma_8(z=0)=0.7913$. The initial conditions were generated at $z=19$ and $z=49$ for simulations A and B respectively, by displacing the particles according to the second-order Lagrangian PT from their initial grid points. The initial  power spectrum of the density fluctuations was computed by CAMB \cite{CAMB}. Taking only the gravitational interaction into account, the simulation was performed with GADGET-2 code \citep{springel05}.

As we estimate the error of the bispectrum from its dispersion among different realizations (see Eq. \ref{sims_variance} in Appendix \ref{Appendix_A} for details), and given that we only have 3 simulations of type B, we divide each of these 3 boxes into 8 sub-boxes. 
Each of these 24 sub-boxes is then treated as if it were an independent realization with smaller box size, $L_b'=937.5\, h/$Mpc, where each of these sub-boxes contains about $512^3$ particles. 
The measurements of the bispectrum from sub-boxes suffer from two issues: 
a) the measurements are not completely independent and more importantly 
b) the sub-boxes are affected by modes larger than sub-box size.
As a consequence of this, a new source of non-Gaussian errors arises for the power spectrum and bispectrum estimation, called beat-coupling effect \cite{beatcoupling1,beatcoupling2,sefusatti06}. 
However, by using the mean density measured in each sub-box instead of the global mean density for the normalization of the density contrast, $\delta\equiv\rho/\bar{\rho}-1$, this effect gets strongly suppressed \cite{Roland}. Hence, we expect that on overlapping scales the bispectrum errors estimated from simulation B to be slightly larger than those from A. This is shown in Appendix \ref{Appendix_A}.

In order to obtain the dark matter field from particles we discretize each box of simulation A and each sub-box of simulation B using $512^3$ grid cells. Thus the size of the grid cells is 4.68 Mpc/$h$ in A and 1.83 Mpc/$h$ in B. We assign the particles to the cells using the count-in-cells prescription. 

More details about the estimation of the bispectrum from simulations and the error bars computation are given respectively in Eq. \ref{bispectrum_estimator}  and \ref{sims_variance} in Appendix \ref{Appendix_A}.

\section{Results}

In order to find the best-fit parameters from Eq. \ref{abc_new}, namely $a_i$, we minimize $$\chi^2\propto \sum_i \left[\left(Q_i^{th}-Q_i^{sims}\right)/\sigma_{Q_i}^{sims}\right]^2$$
 using a set of triangle configurations: $k_2/k_1=1.0, 1.5, 2.0, 2.5$ and $\theta_{12}/\pi=0.1, 0.2,\dots, 0.9$, at different redshifts: $z=0, 0.5, 1.0, 1.5$\footnote{ For simulation A the $z$ used are $z=0, 0.5, 1.0, 1.5$ whereas for simulation B $z=0, 0.42, 1.0, 1.5$}. The algorithm used for the minimization is {\it amoeba} \citep{amoeba}. In our analysis we neglect that the errors of the data points are correlated. However, since our errors are small (typically less than 5\%) we expect that error correlations do not play an important role for this method: the dominating source of the error of our fitting formula given in Eq. \ref{abc_new} comes from the imperfection of the functional form of the fitting formula and not from the uncertainties in the simulation data. We have checked that the result converges from different starting points. The resulting best-fit values are shown in Table \ref{tabla_fit}.
 
\begin{table}
\begin{center}
\begin{tabular}{lll}
$a_1=0.484$ & $a_2=\,3.740$ & $a_3=-0.849$ \\
$a_4=0.392$ & $a_5=\,1.013$ & $a_6=-0.575$ \\
$a_7=0.128$ & $a_8=-0.722$ & $a_9=-0.926$ 
\end{tabular}
\end{center}
\caption{Best-fit parameters (according to Eq. \ref{abc_new}) derived by combining data from simulations A and B, using different triangle configurations $\theta_{12}/\pi=0.1, 0.2,\dots,0.9$ and $k_2/k_1=1.0,1.5,2.0,2.5$ and at different redshifts $z=0,0.5,1.0,1.5$.}
\label{tabla_fit}
\end{table}

We have also checked that this is a very good fit not only for the reduced bispectrum $Q$ but also for the bispectrum $B$.

\begin{figure}
\centering
\includegraphics[clip=false, trim= 110mm 35mm 45mm 35mm,scale=0.28]{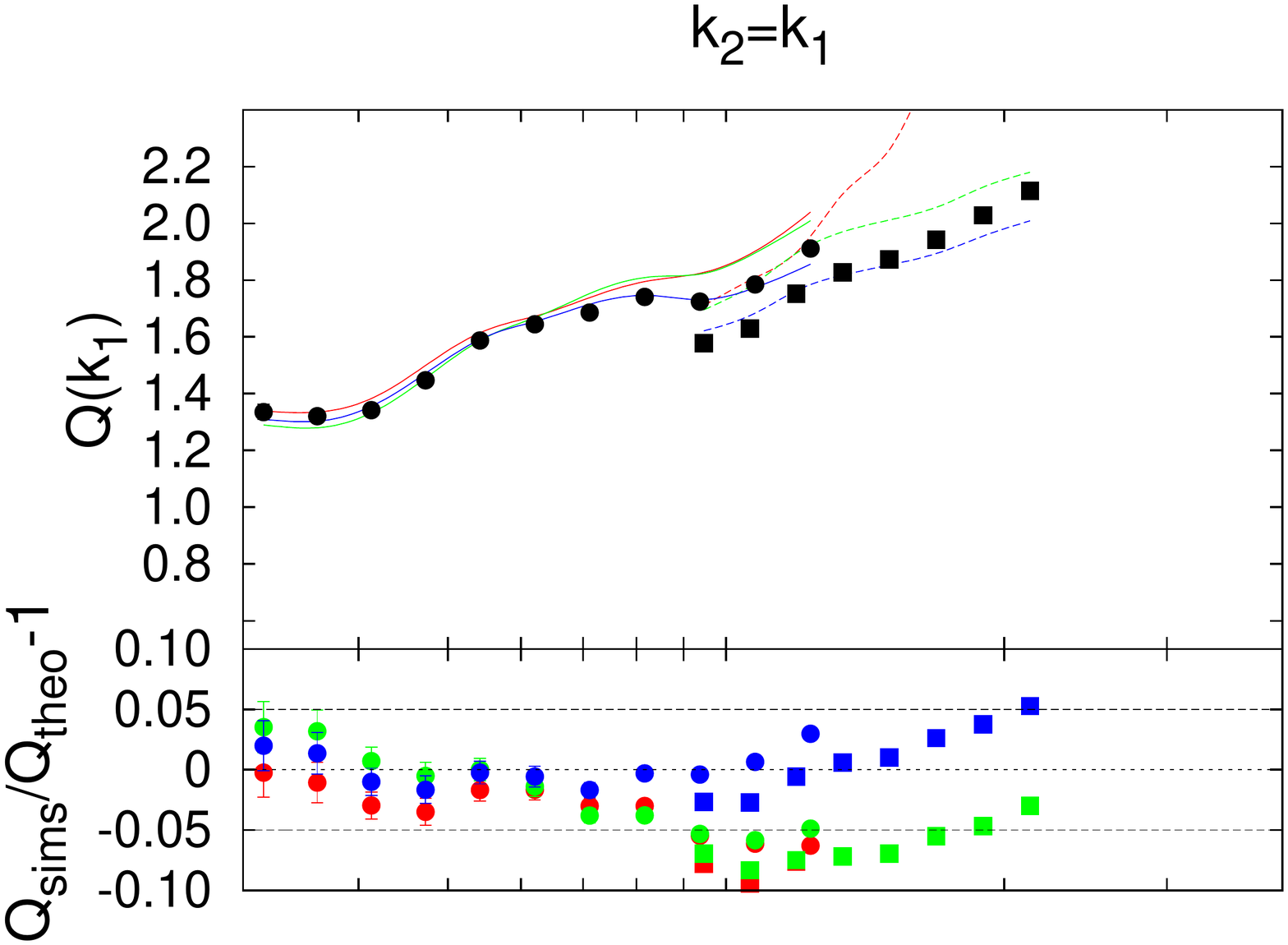}
\includegraphics[clip=false,trim= 45mm 35mm 45mm 35mm, scale=0.28]{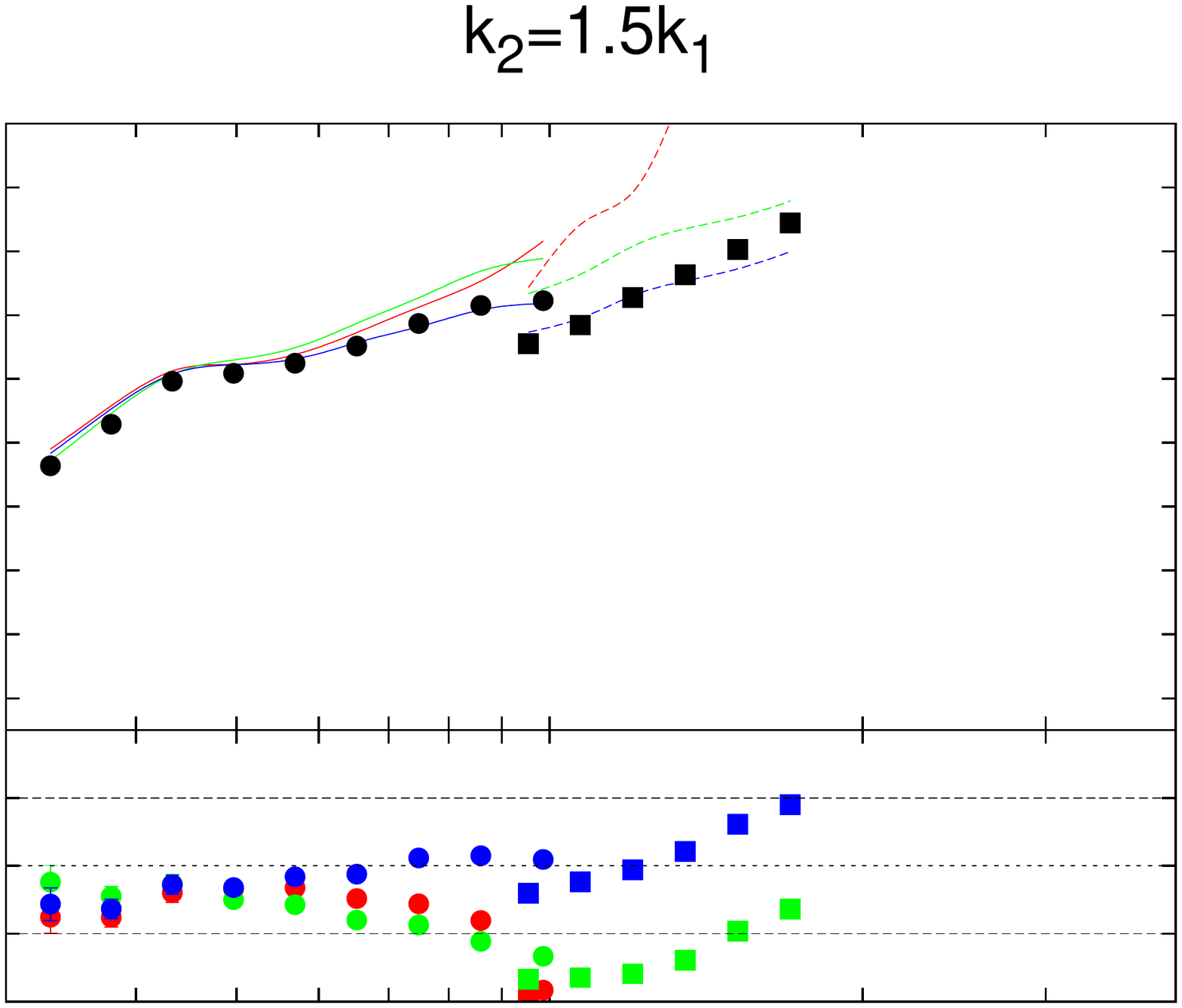}
\includegraphics[clip=false,trim= 45mm 35mm 110mm 35mm, scale=0.28]{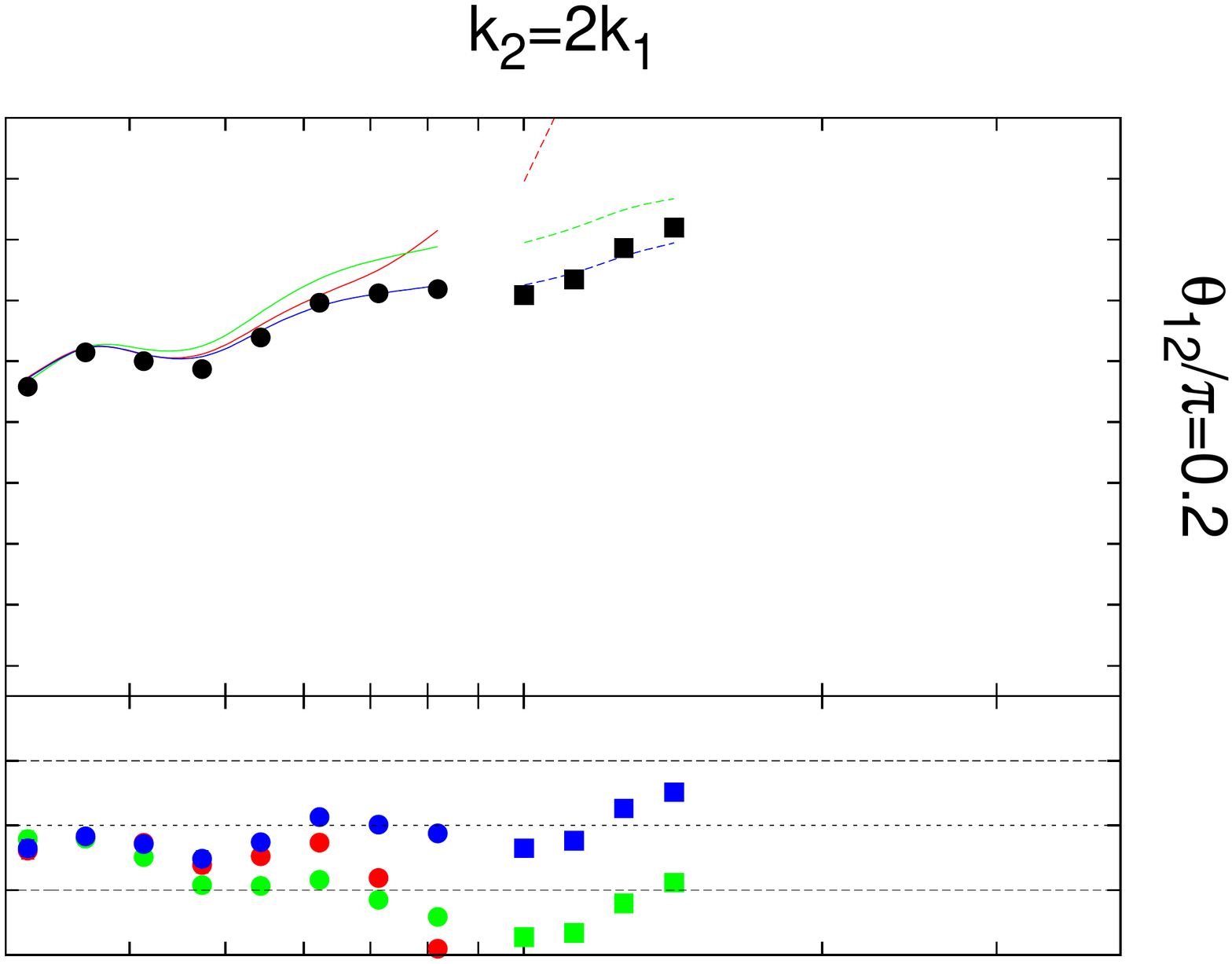}

\includegraphics[clip=false, trim= 110mm 35mm 45mm 35mm,scale=0.28]{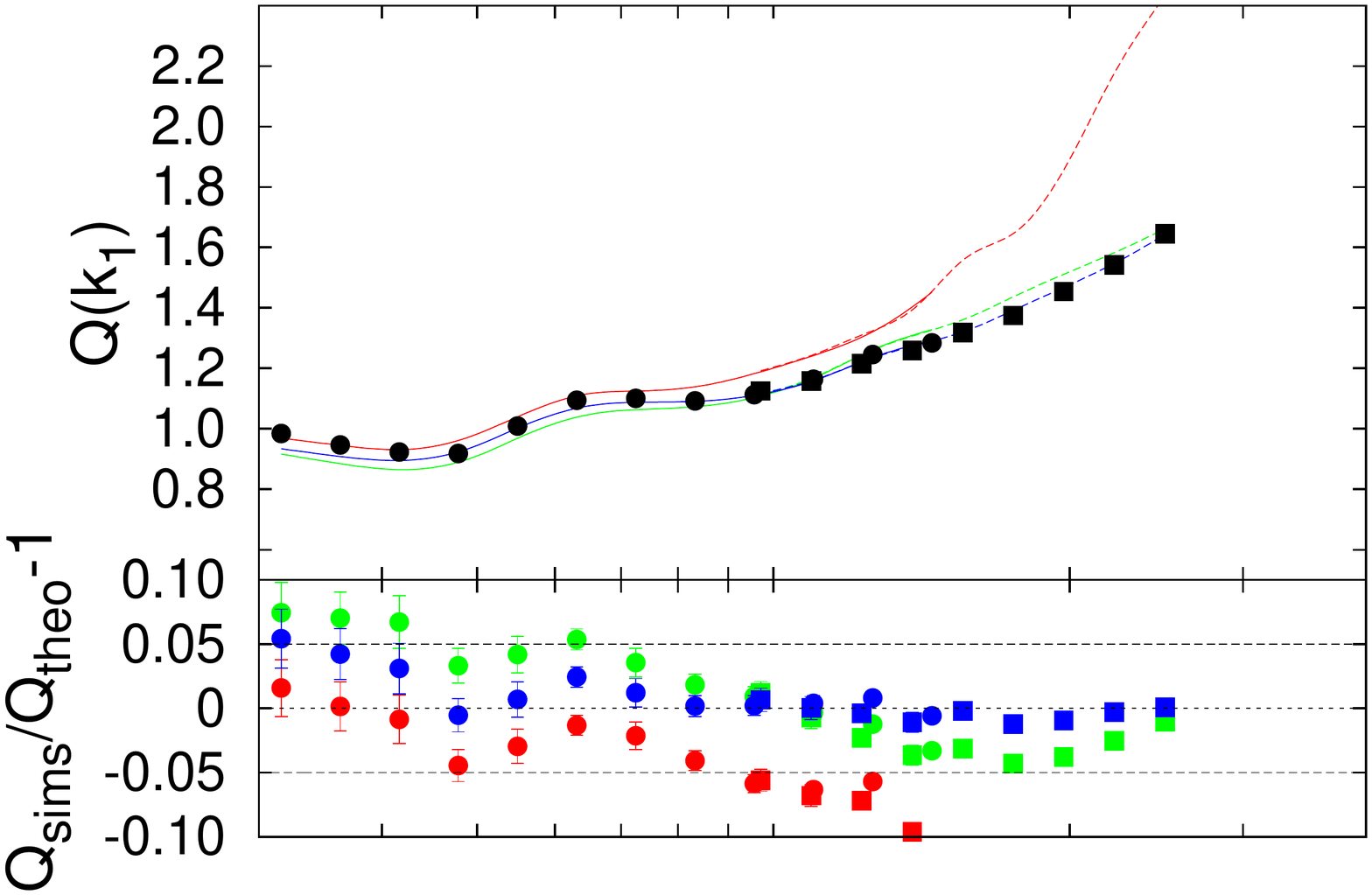}
\includegraphics[clip=false,trim= 45mm 35mm 45mm 35mm, scale=0.28]{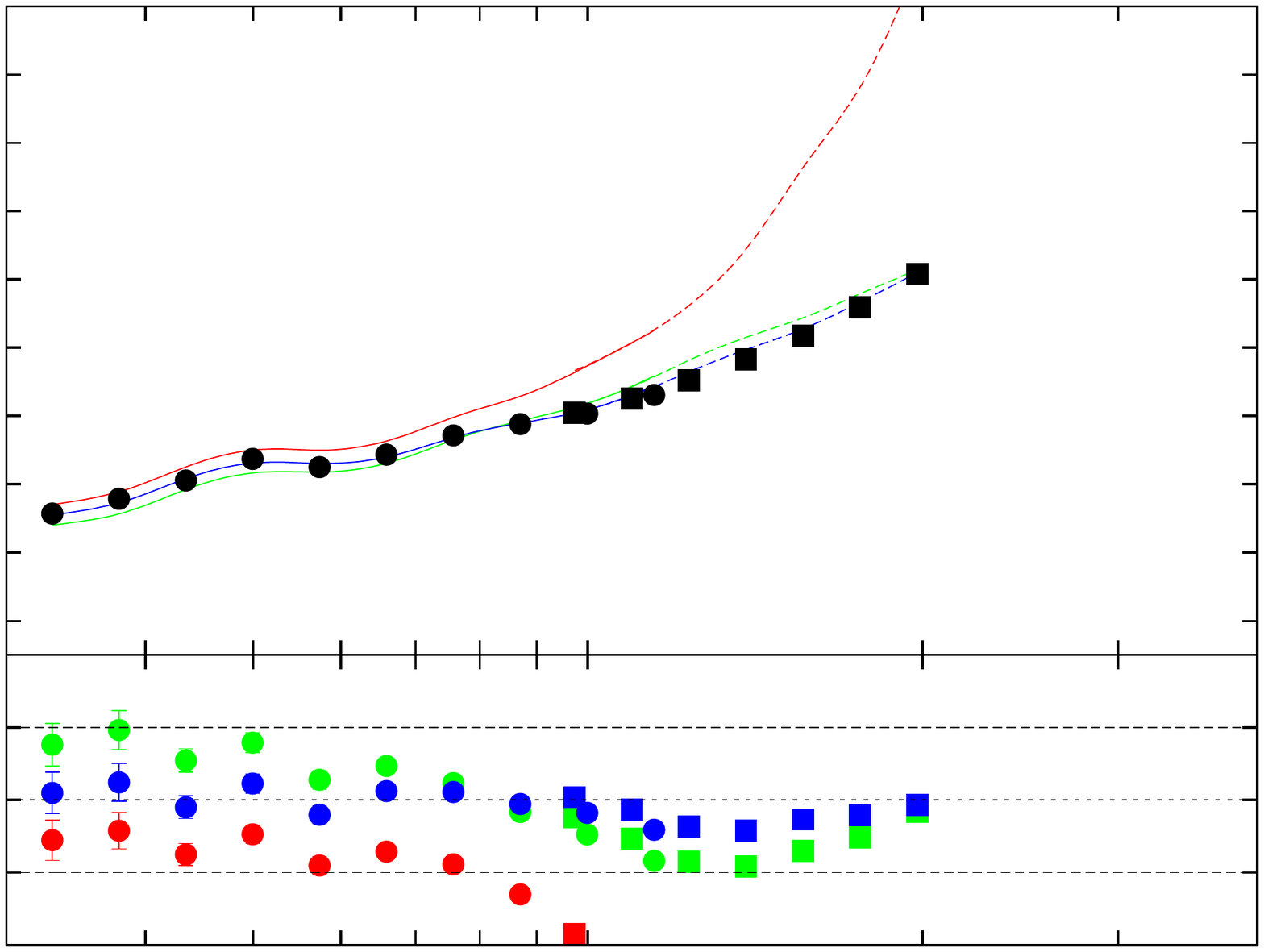}
\includegraphics[clip=false,trim= 45mm 35mm 110mm 35mm, scale=0.28]{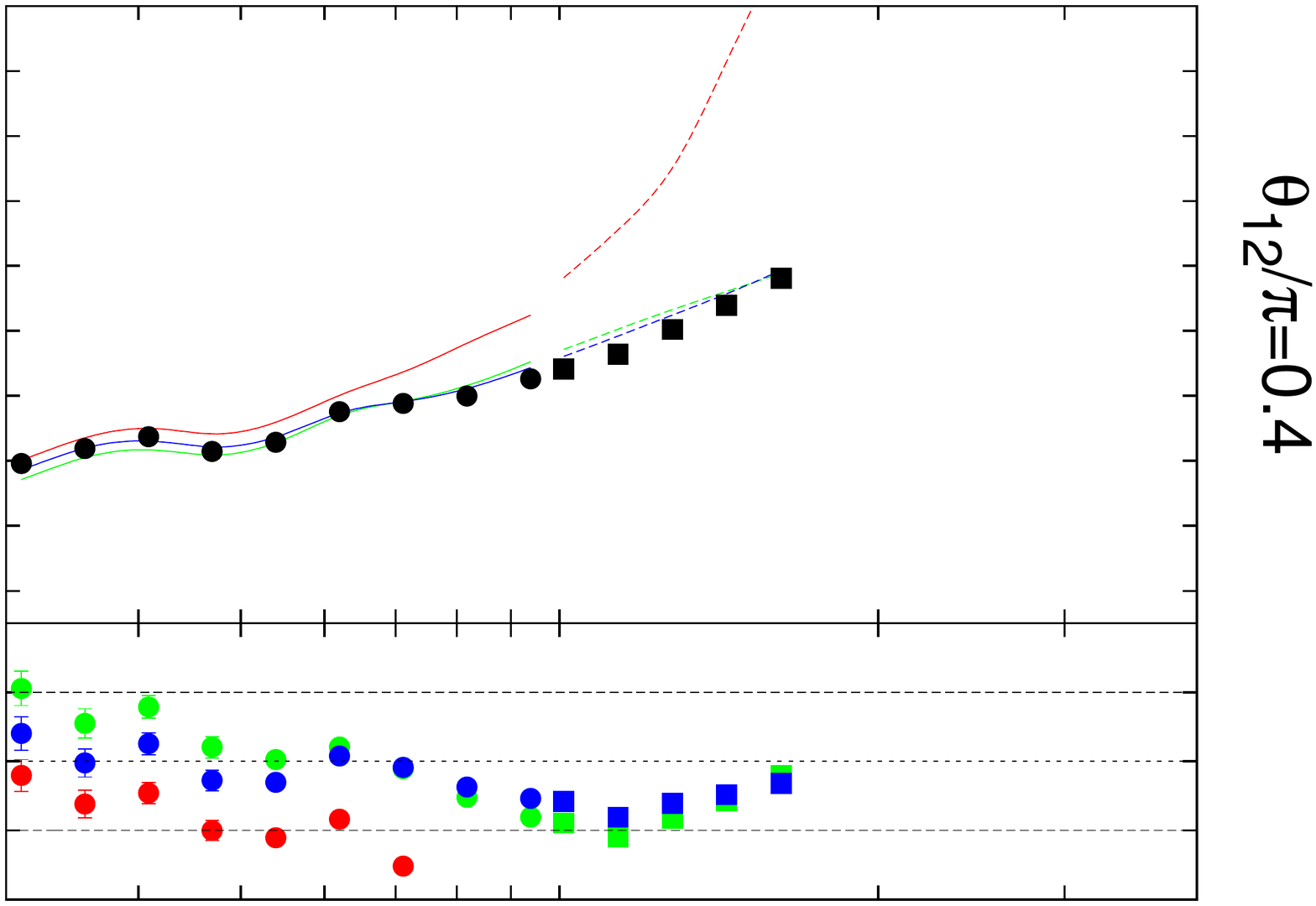}

\includegraphics[clip=false, trim= 110mm 35mm 45mm 35mm,scale=0.28]{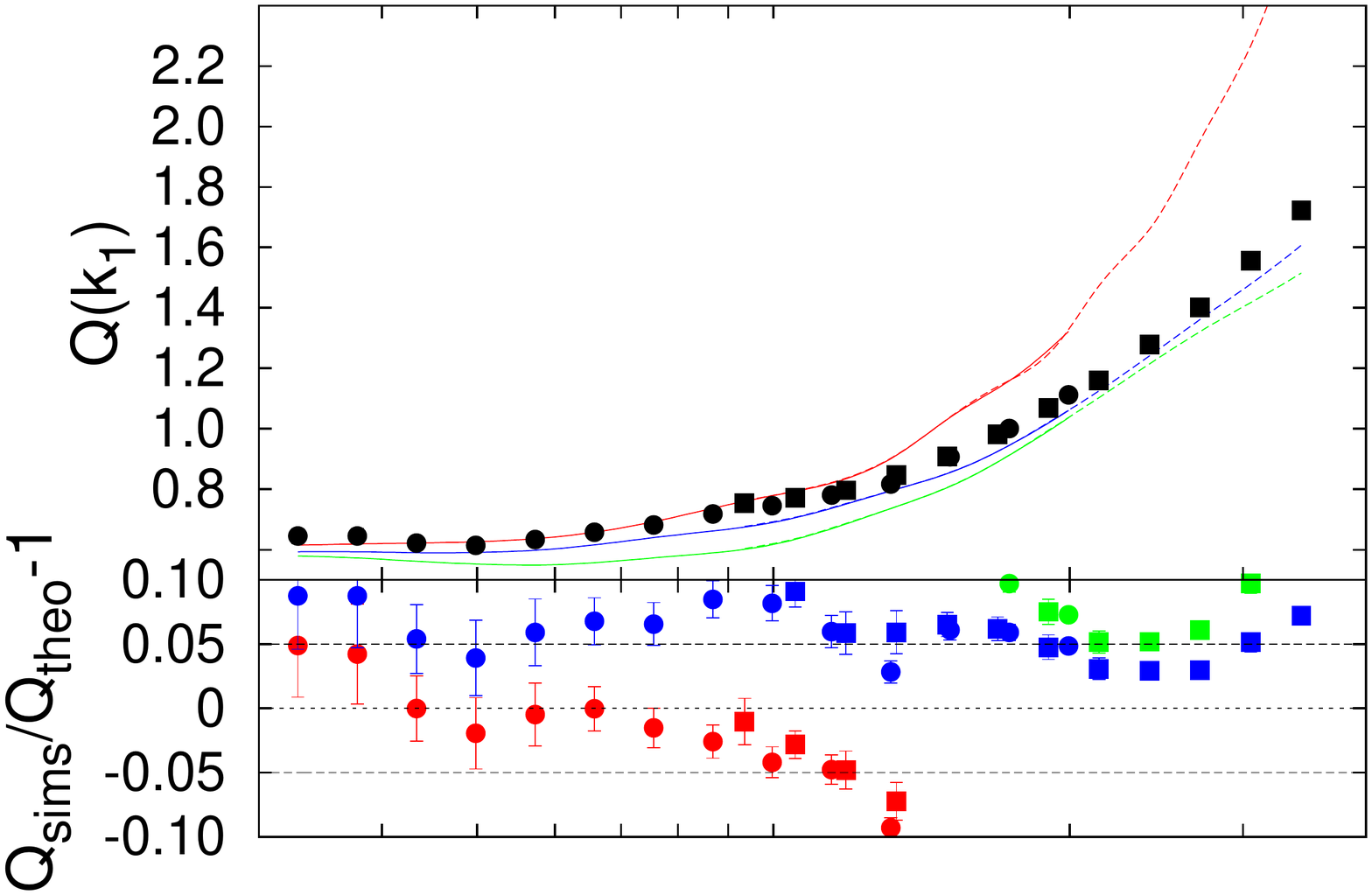}
\includegraphics[clip=false,trim= 45mm 35mm 45mm 35mm, scale=0.28]{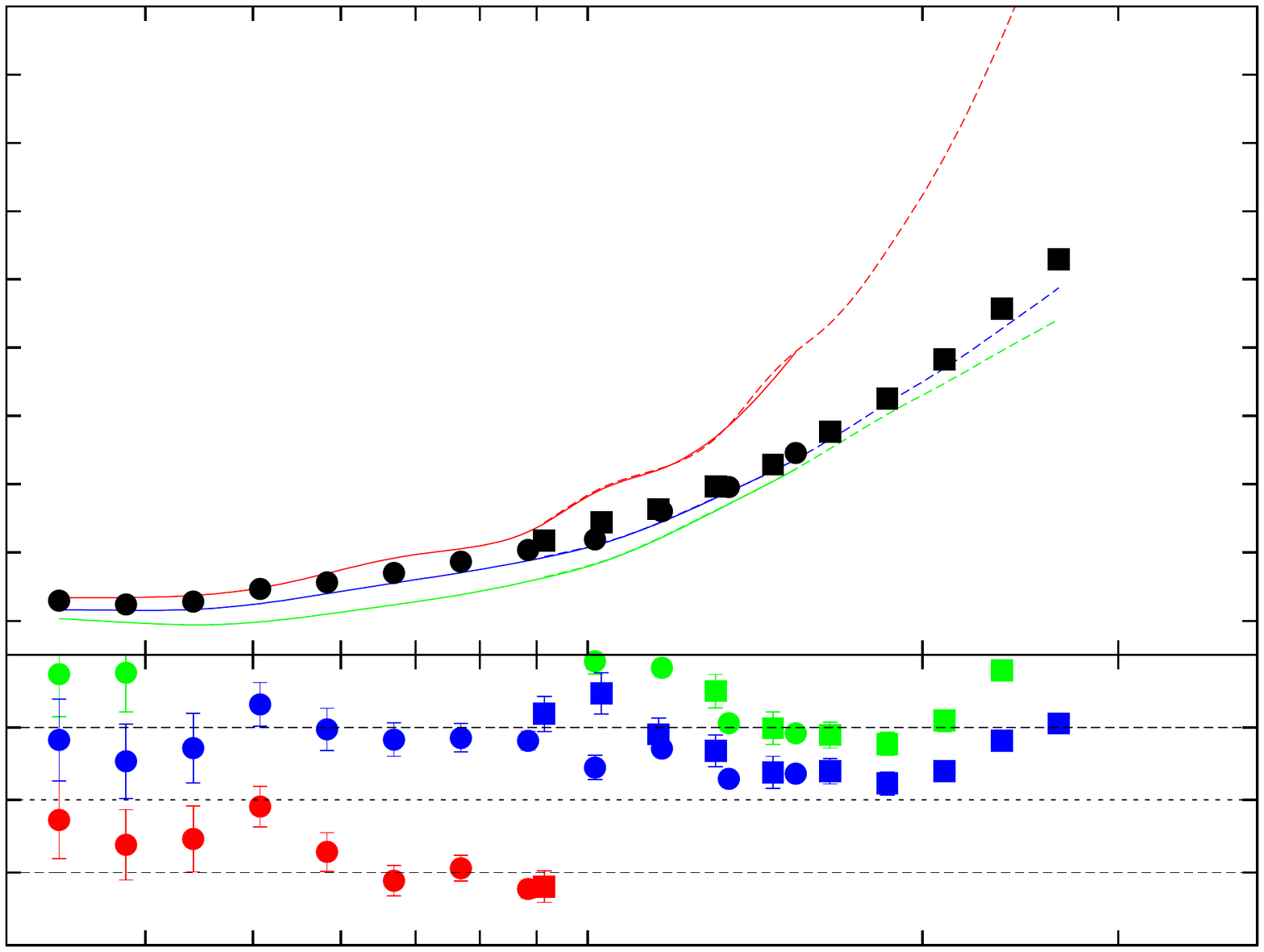}
\includegraphics[clip=false,trim= 45mm 35mm 110mm 35mm, scale=0.28]{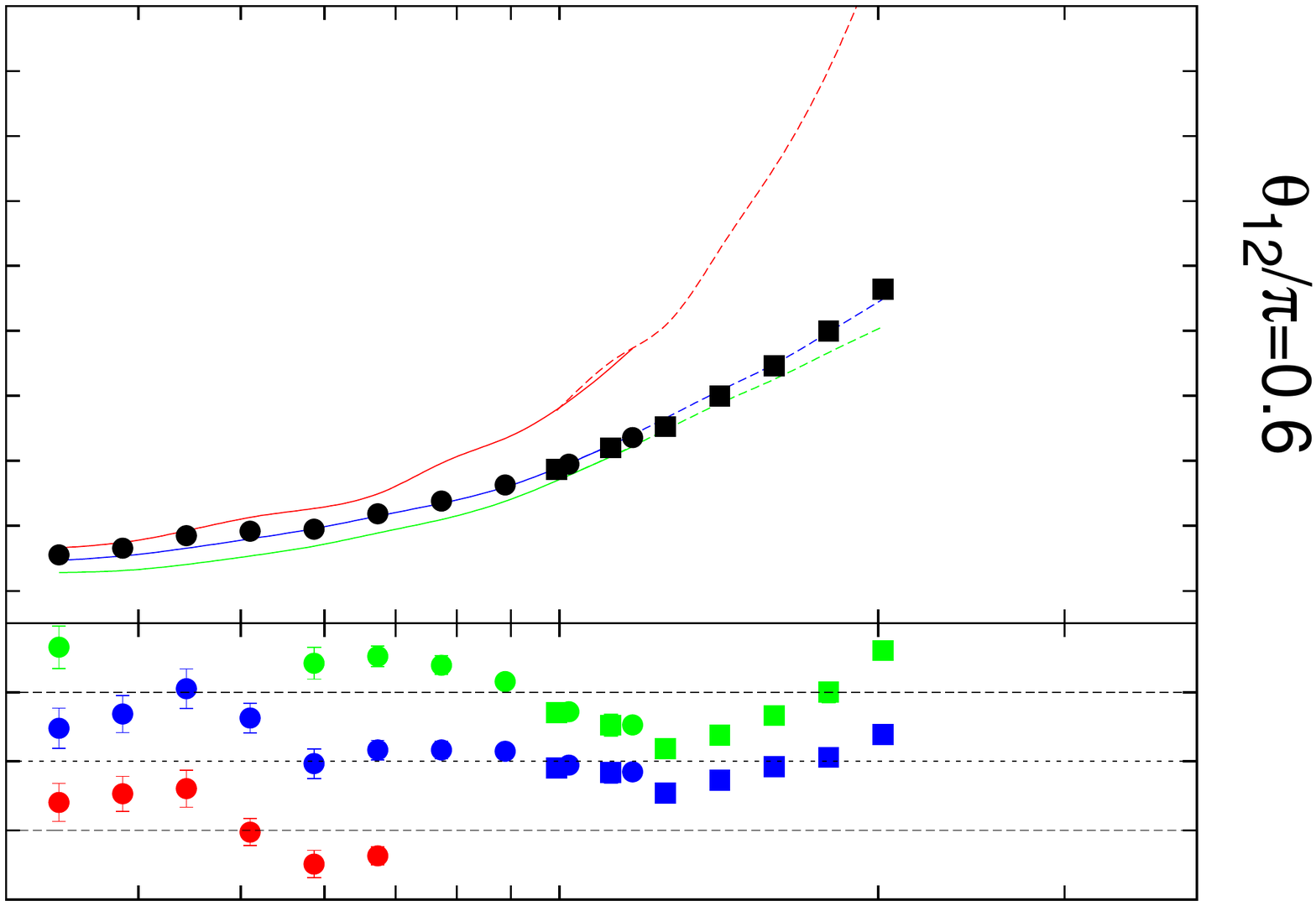}

\includegraphics[clip=false, trim= 110mm 30mm 45mm 35mm,scale=0.28]{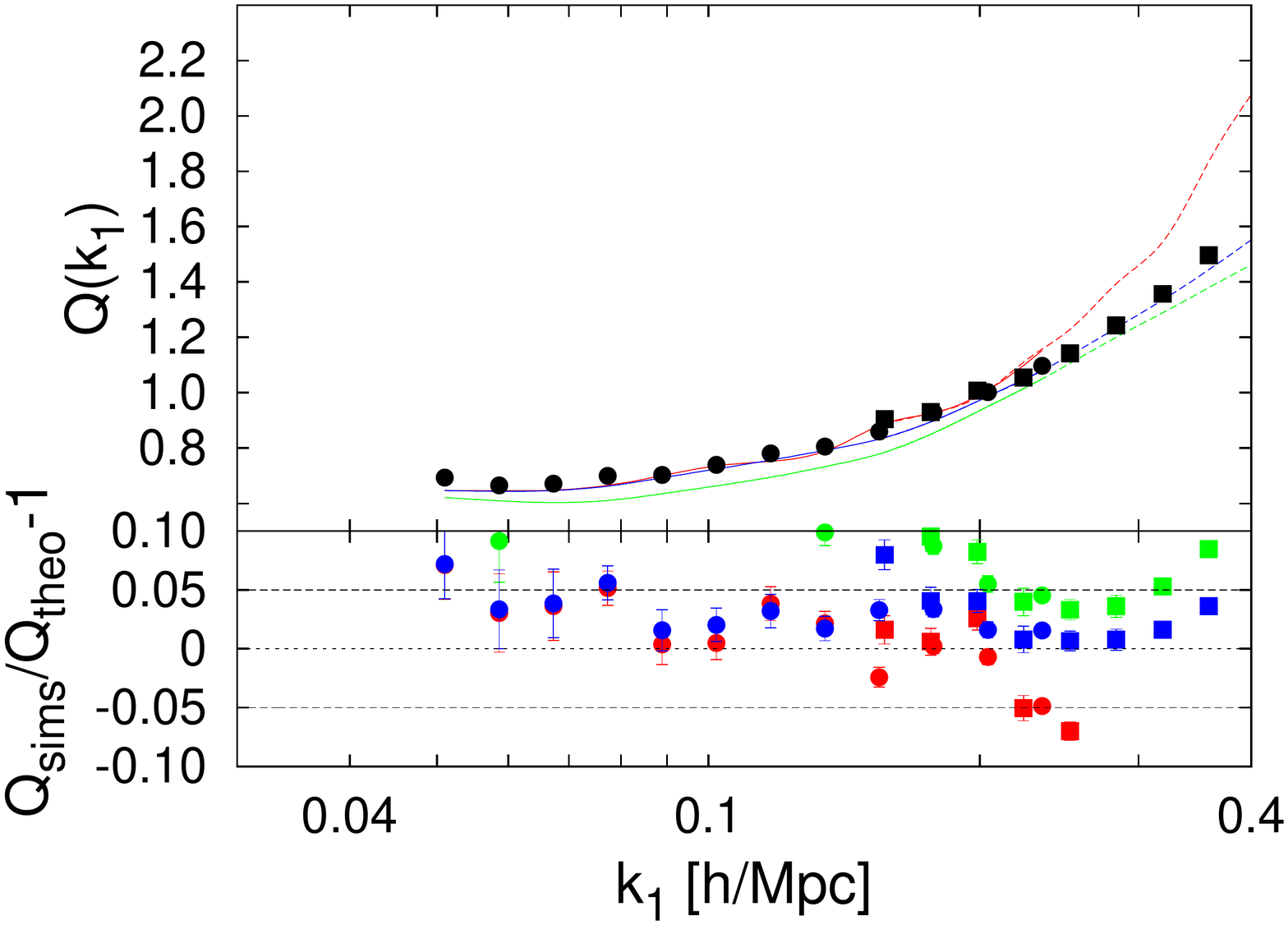}
\includegraphics[clip=false,trim= 45mm 30mm 45mm 35mm, scale=0.28]{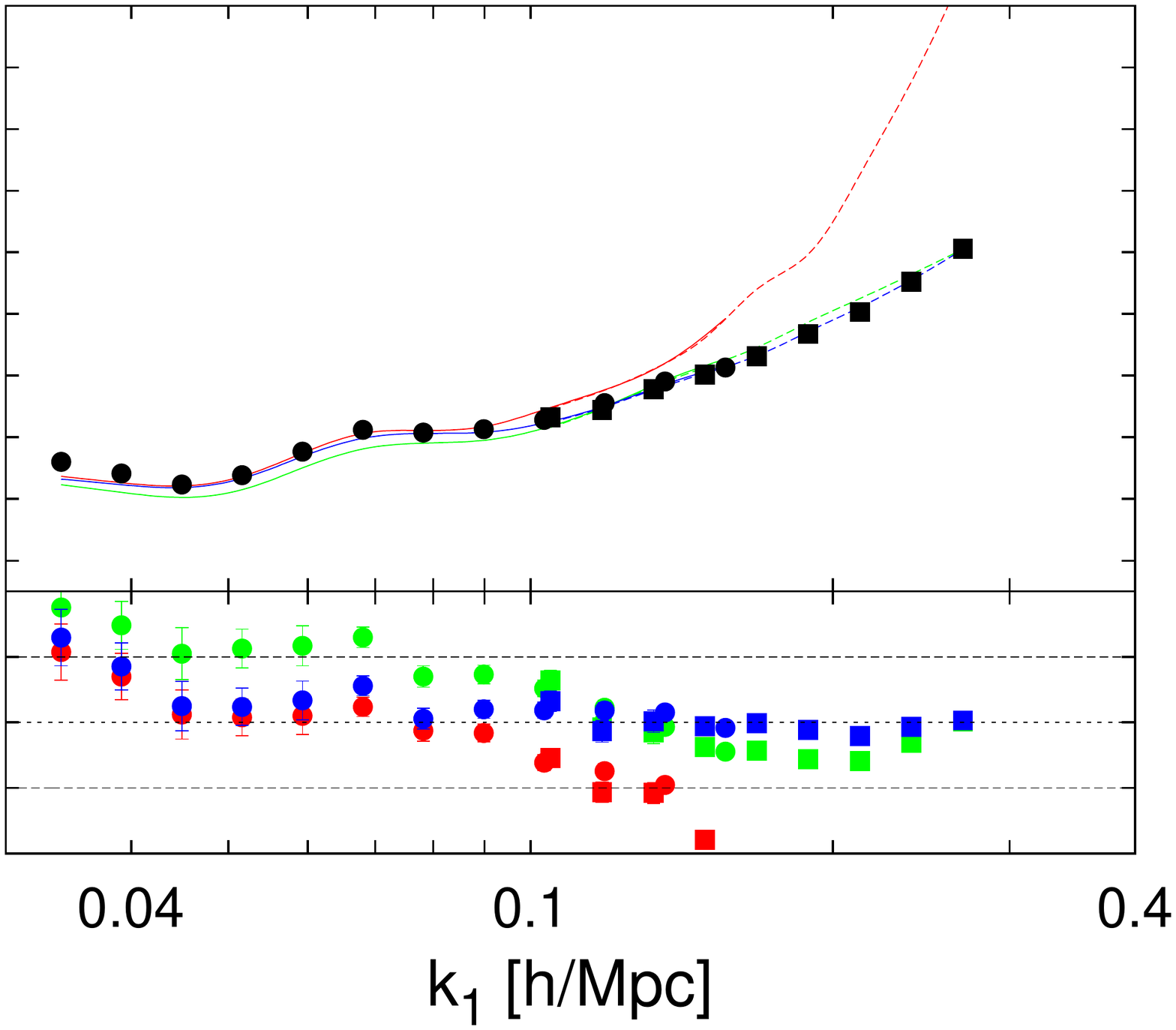}
\includegraphics[clip=false,trim= 45mm 30mm 110mm 35mm, scale=0.28]{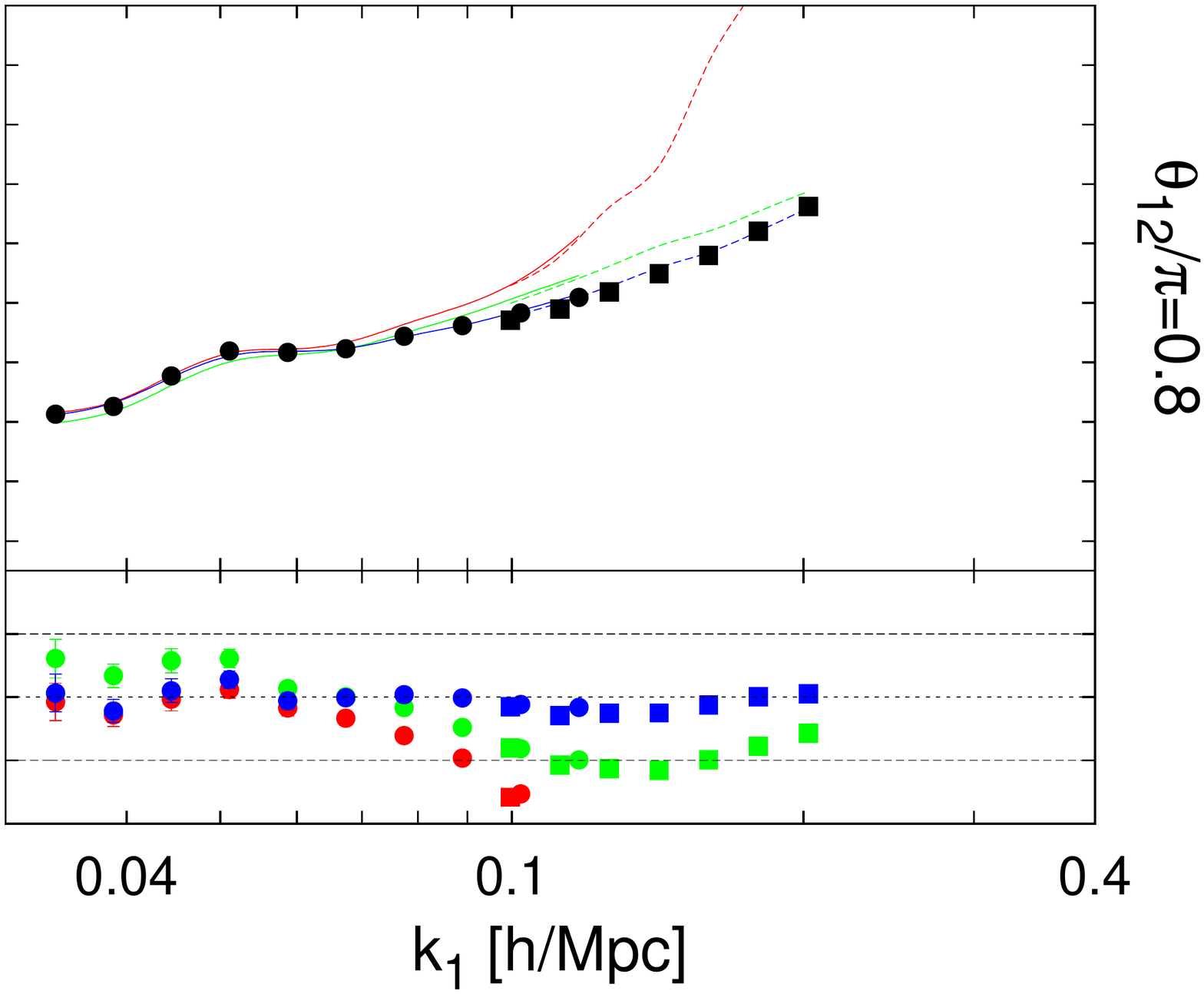}

\caption{In the {\it main} panels: $Q$ vs. $k_1$ for N-body simulation (black circles for simulation A and black squares for simulation B), for 1-loop correction (red solid lines to fit simulation A data and dashed line to fit simulation B), for SC formula (green solid line to fit A and green dashed line to fit B) and for our model (blue solid line to fit A and blue dashed line to fit B) at $z=0$. In the {\it sub-panels}, the ratio between simulations and different theoretical models is shown: 1-loop (red points), SC (green points) and our model (blue points) for different triangles configurations. Circle symbols comes from simulation A and squares symbol from simulation B. The error bars show the measured errors from the simulations.
Dashed lines mark 5\% deviation from data. From left to right panels: $k_2/k_1=1.0, 1.5, 2.0$. From top to bottom panels: $\theta_{12}/\pi=0.2, 0.4, 0.6, 0.8$. }

\label{Qz0}

\end{figure}

\begin{figure}
\centering
\includegraphics[clip=false, trim= 110mm 35mm 45mm 35mm,scale=0.28]{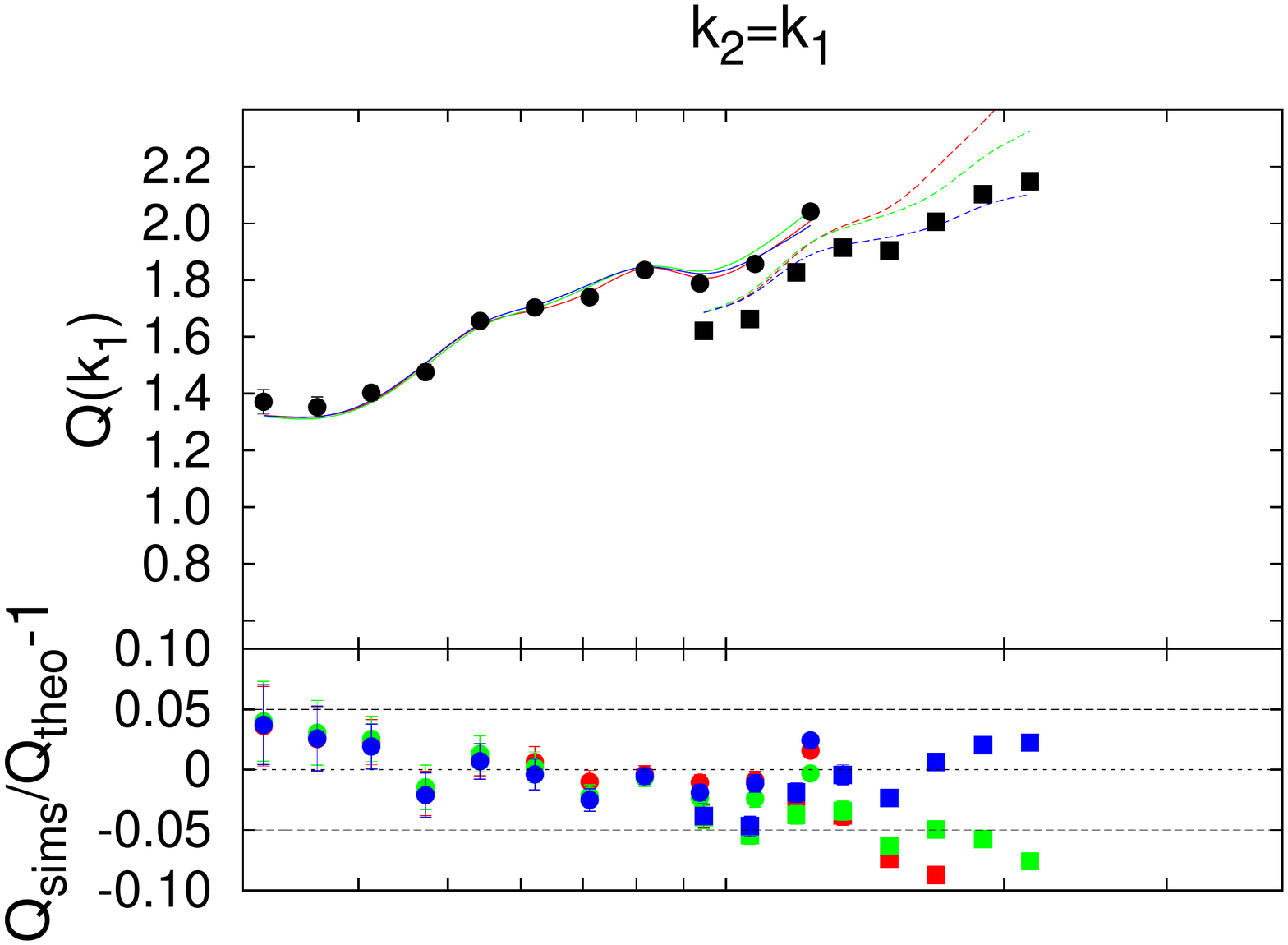}
\includegraphics[clip=false,trim= 45mm 35mm 45mm 35mm, scale=0.28]{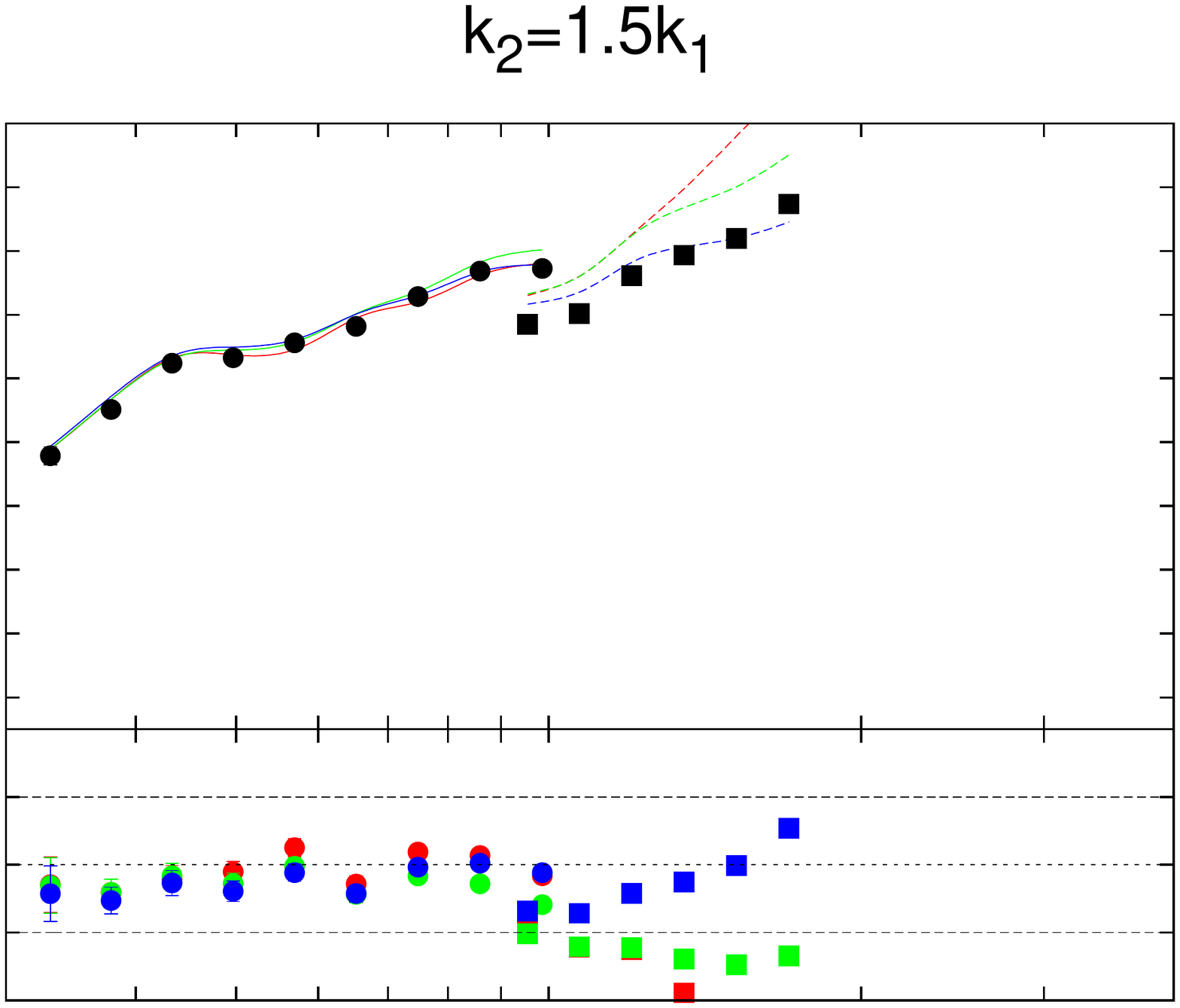}
\includegraphics[clip=false,trim= 45mm 35mm 110mm 35mm, scale=0.28]{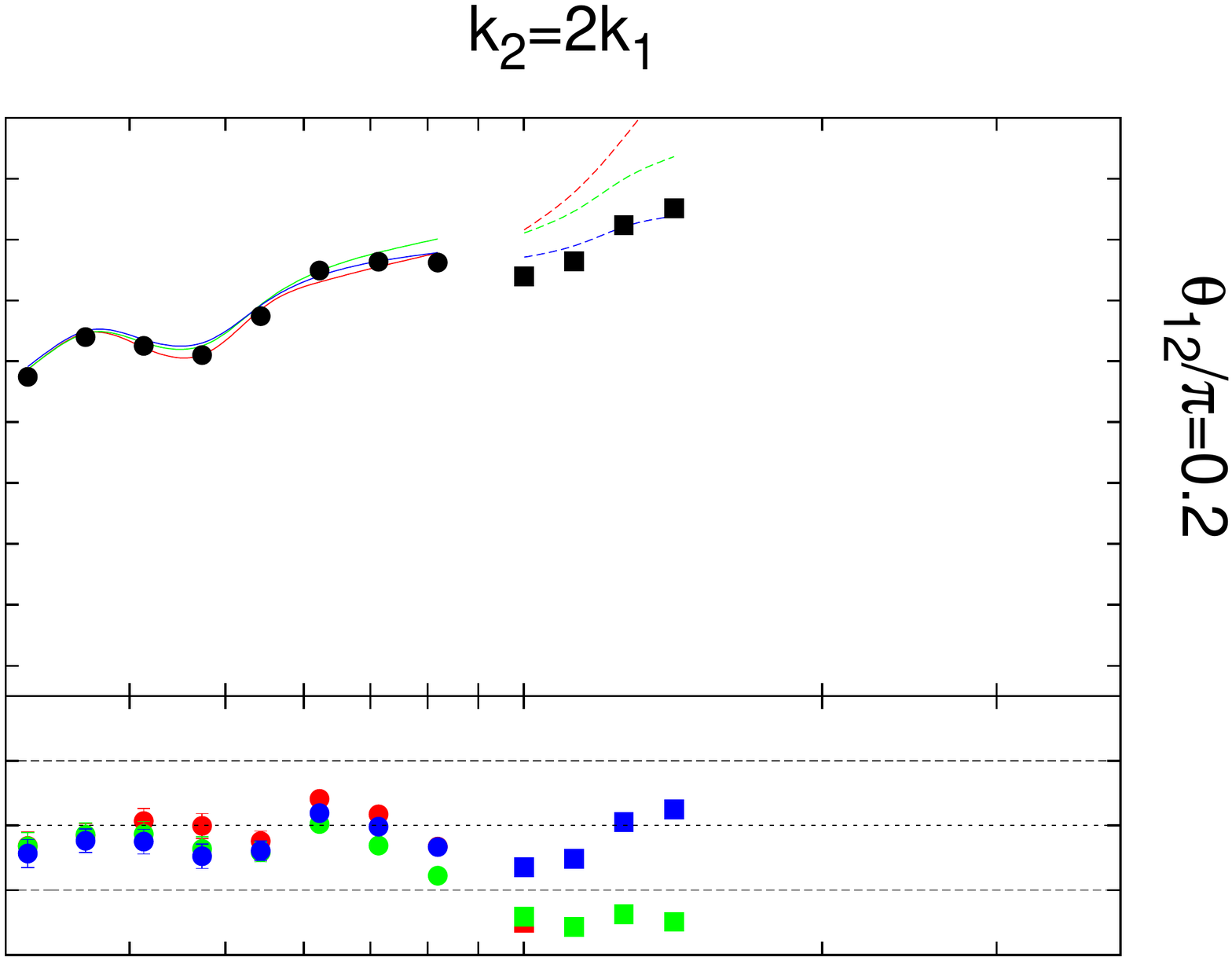}

\includegraphics[clip=false, trim= 110mm 35mm 45mm 35mm,scale=0.28]{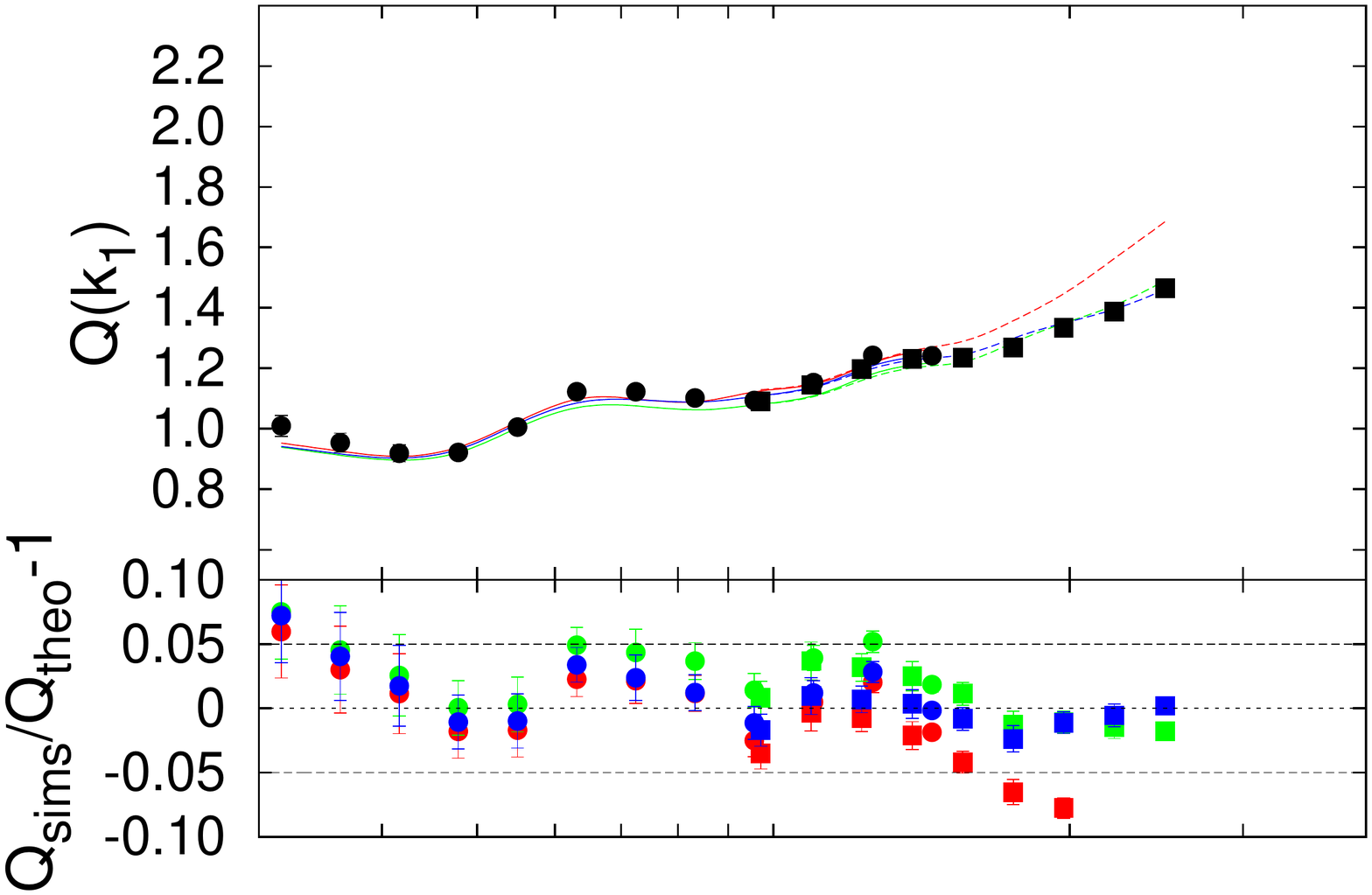}
\includegraphics[clip=false,trim= 45mm 35mm 45mm 35mm, scale=0.28]{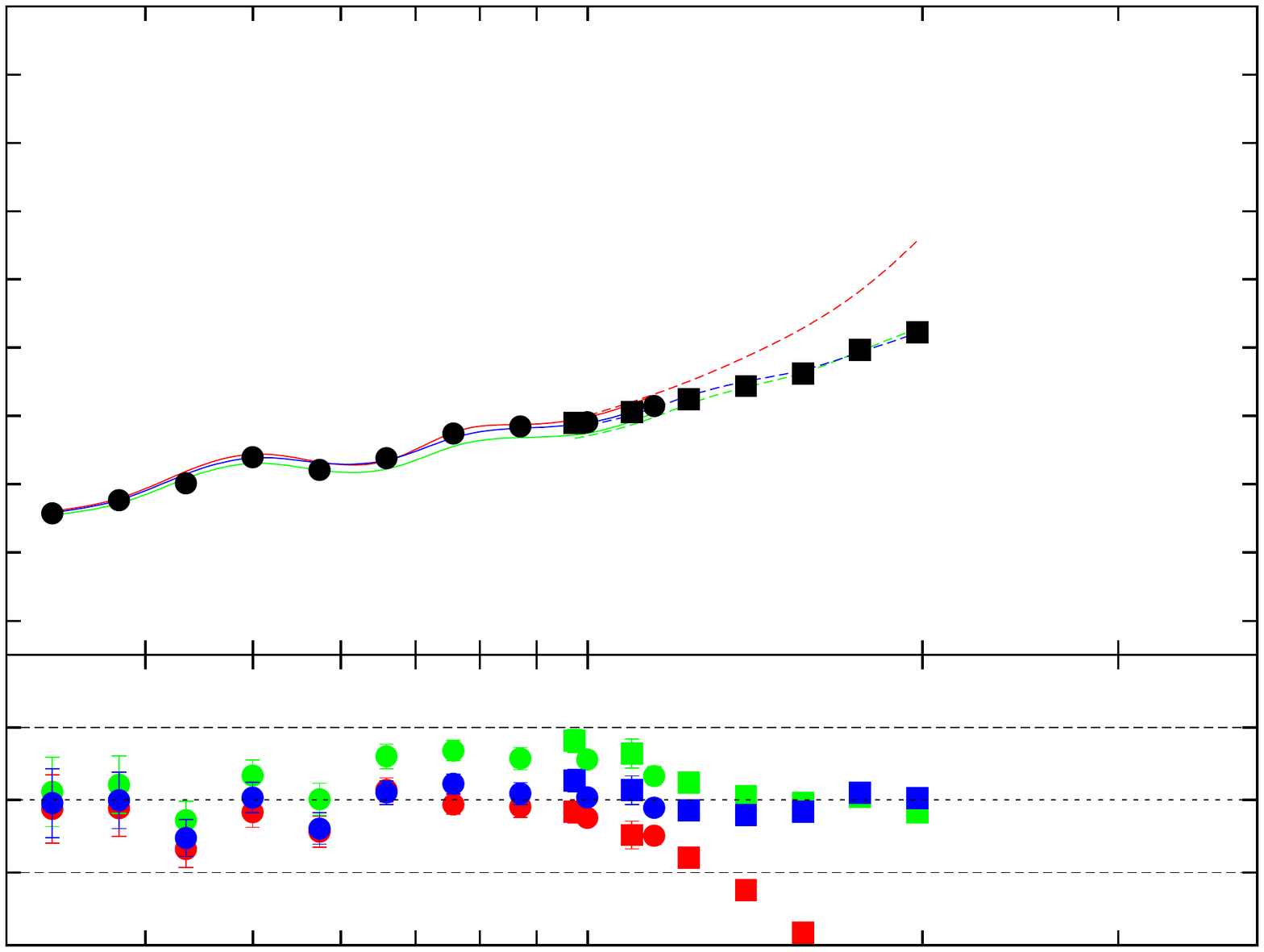}
\includegraphics[clip=false,trim= 45mm 35mm 110mm 35mm, scale=0.28]{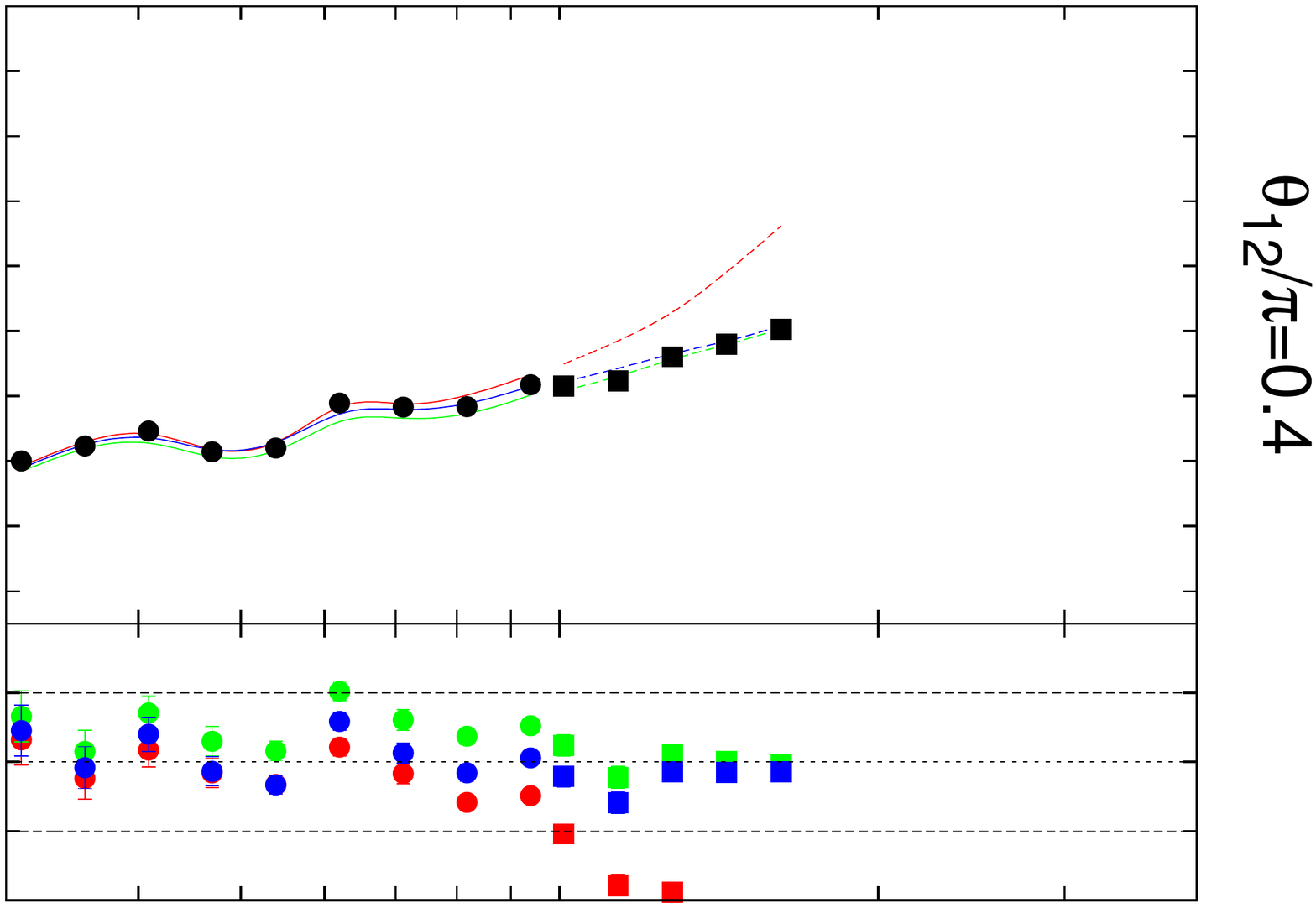}

\includegraphics[clip=false, trim= 110mm 35mm 45mm 35mm,scale=0.28]{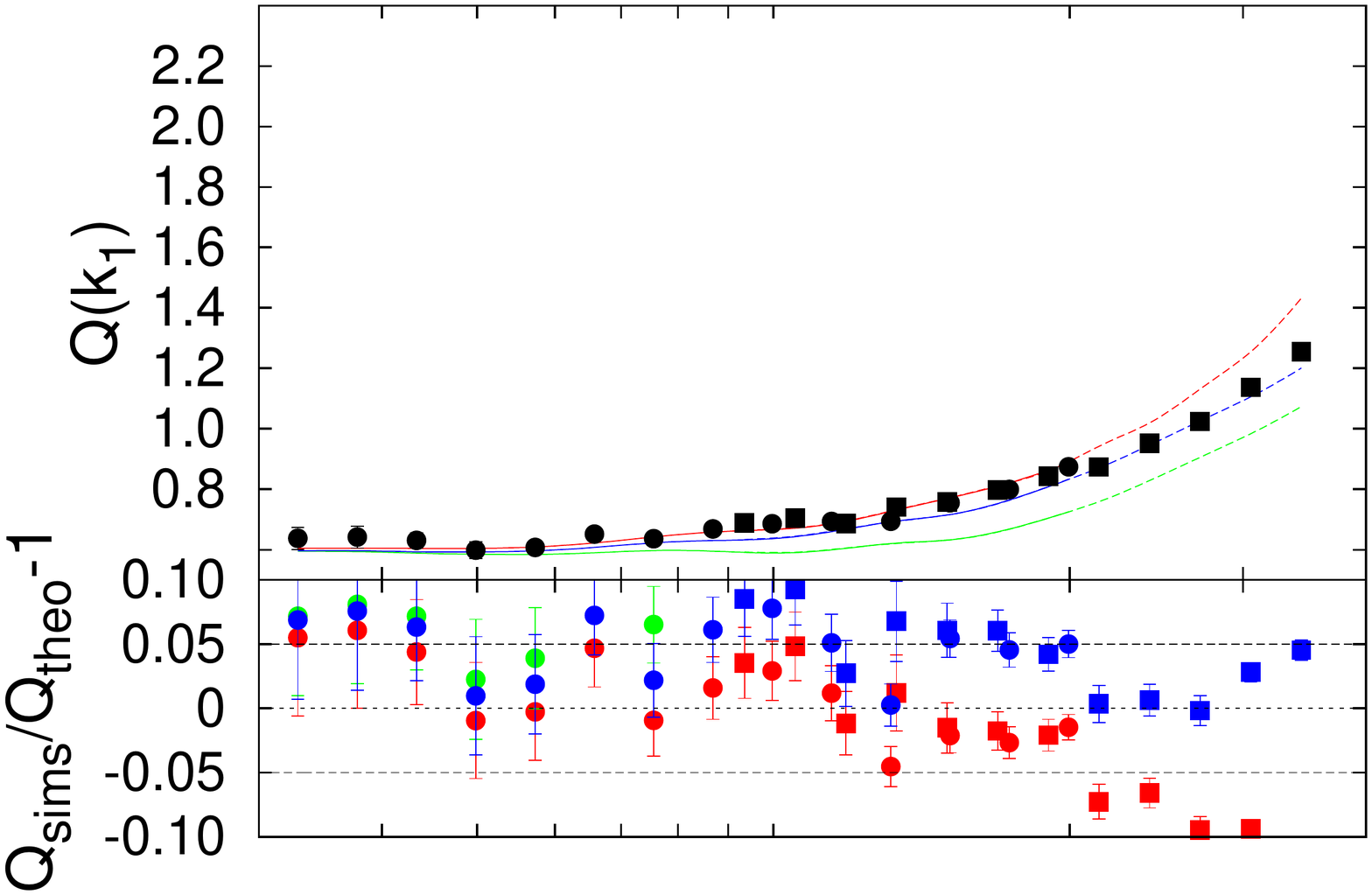}
\includegraphics[clip=false,trim= 45mm 35mm 45mm 35mm, scale=0.28]{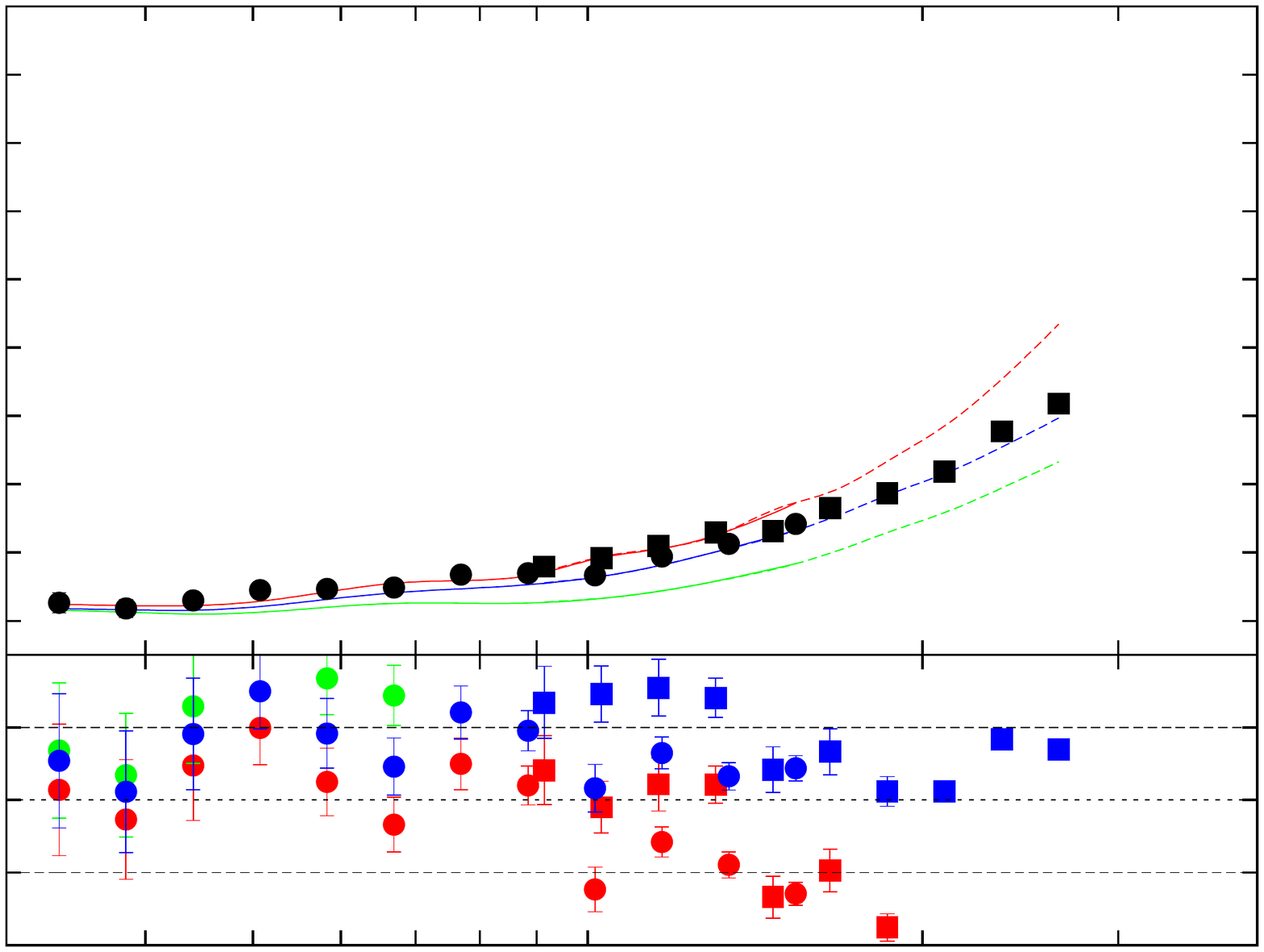}
\includegraphics[clip=false,trim= 45mm 35mm 110mm 35mm, scale=0.28]{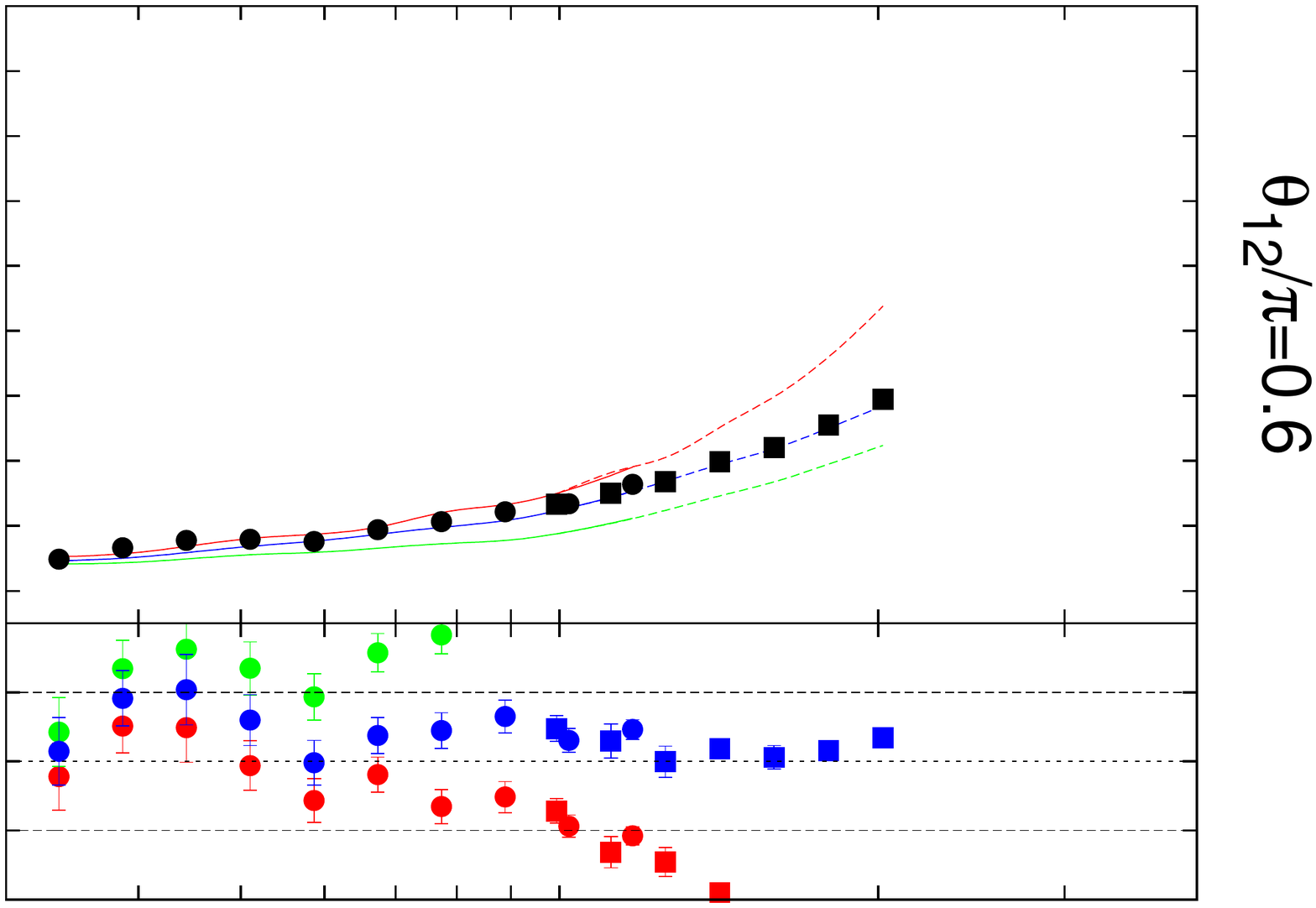}

\includegraphics[clip=false, trim= 110mm 30mm 45mm 35mm,scale=0.28]{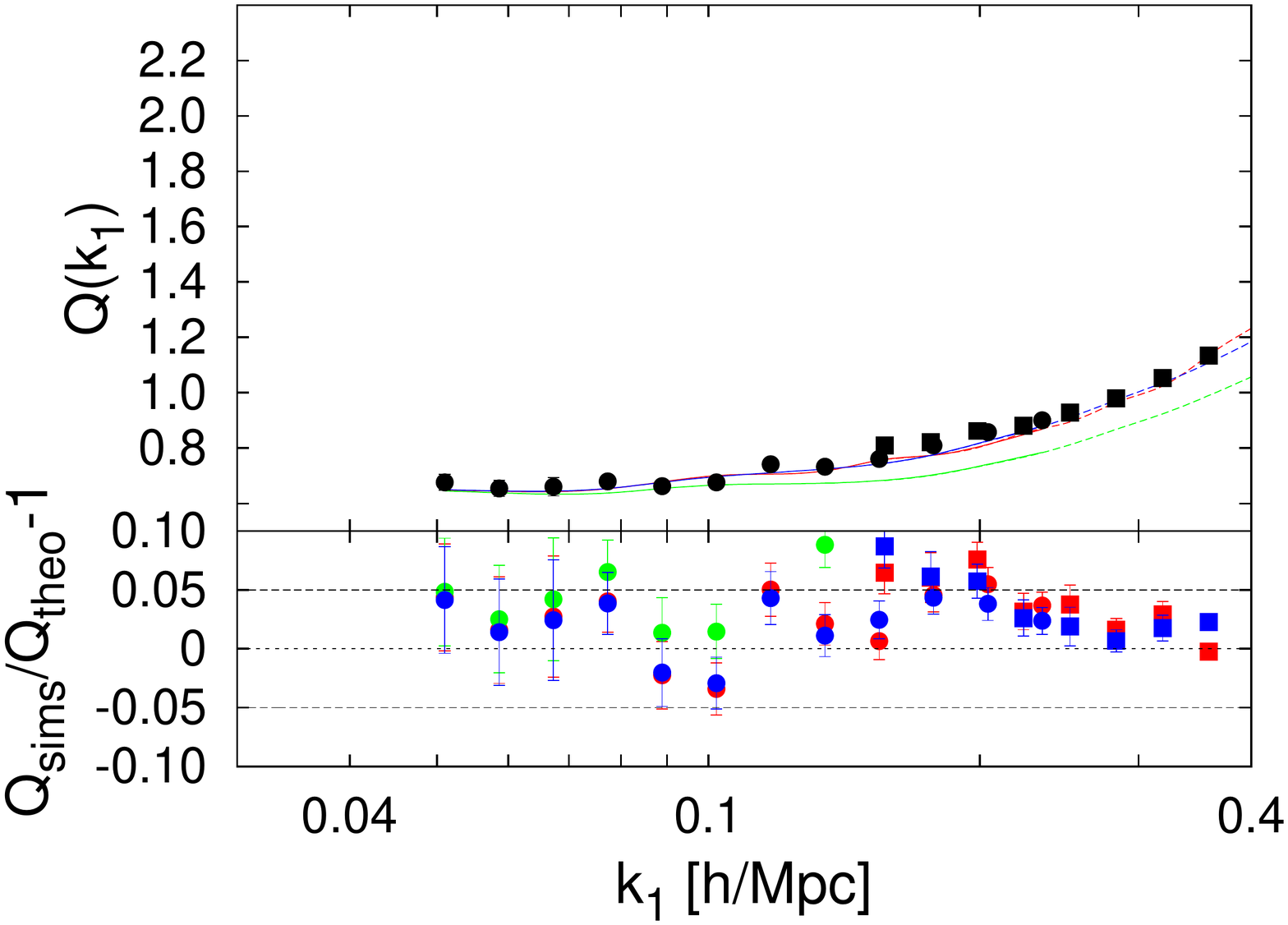}
\includegraphics[clip=false,trim= 45mm 30mm 45mm 35mm, scale=0.28]{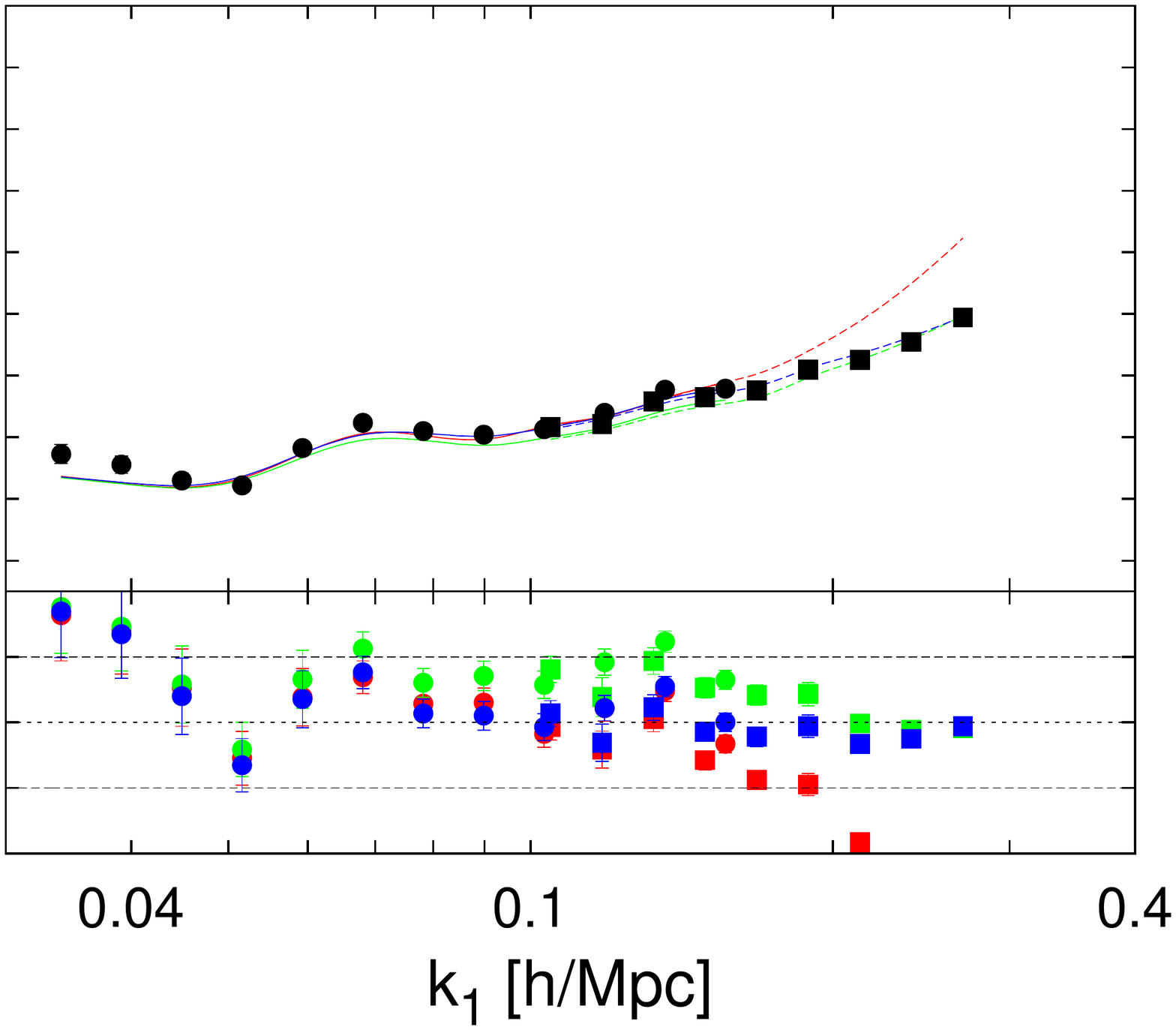}
\includegraphics[clip=false,trim= 45mm 30mm 110mm 35mm, scale=0.28]{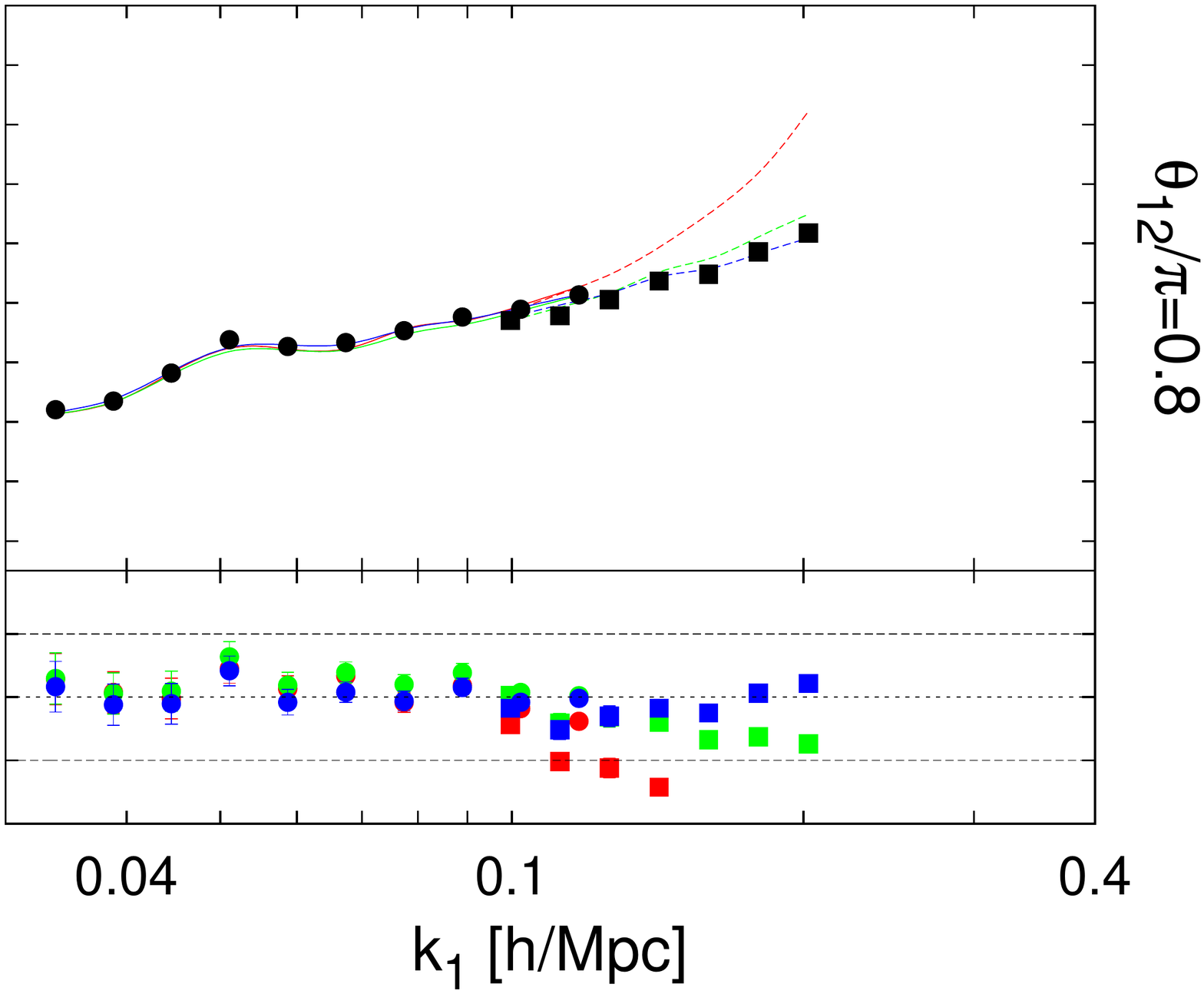}

\caption{  $Q(k_1)$ for different triangles configurations at $z=1$. Same notation that in Fig. \ref{Qz0}.}

\label{Qz1}

\end{figure}

 In Fig. \ref{Qz0} and \ref{Qz1} we show the results of our fit and also the predictions of two other models for different triangles configurations for $z=0$ (Fig. \ref{Qz0}) and for $z=1$ (Fig. \ref{Qz1}). These two models are 1-loop Eulerian PT (see Appendix \ref{Appendix_C}) and the SC method (Eq. \ref{SC_formula}) + smoothed-$n$. In each plot we show the reduced bispectrum $Q$ vs. $k_1$ for: N-body data (black circles for simulations A and black squares for simulations B), 1-loop correction (red solid line corresponding to data of simulations A and red dashed line to data  of simulations B), SC formula (green solid line for A and green dashed line for B) and our model (blue solid and dashed line for A and B respectively). In the bottom part of each panel we show the deviation of  these models with respect to the N-body data: red symbols depict the deviation of the 1-loop prediction with respect to the data, green symbols are the deviation of the SC formula and blue symbols are the deviation of our model. Circles are the deviation with respect to simulation A and squares with respect to simulation B. 
The error bars show the error of the bispectrum measured from the simulations (see Appendix \ref{Appendix_A} for details).
Each panel shows a different triangle configuration: from left to right $k_2/k_1=1.0,\, 1.5,\, 2.0$ and from top to bottom $\theta_{12}/\pi=0.2,\, 0.4,\, 0.6,\, 0.8$.
 In order to avoid sample variance effects, we only use data points with $k_i>0.03$ $h$/Mpc for simulation A and $k_i>0.09$ $h$/Mpc for simulation B.
 
 For some triangle configurations, we observe a mismatch between the data points in the overlapping region of simulations A and B . This is due to the fact that the data points do not exactly correspond to the same triangle configurations. The central value for $k_i$ is the same for both simulations in each panel. However, since A and B have different fundamental frequencies $k_f\equiv2\pi/L_b$ and we take the bin width to be $\Delta k=3k_f$, the $k$-space over which we average the bispectrum for each bin is different (see Appendix \ref{Appendix_A} for further explanation). 
 Hence, when one computes the effective $\tilde k_i$ using Eq. \ref{k_correction} one obtains that, especially for elongated triangle configurations, the simulations do not represent the same triangle configuration. Since the theoretical predictions (both 1-loop, SC and our model) are computed from $\tilde{k_i}$ there is also a mismatching between solid and dashed lines for the same reason. This effect is also noted in \cite{sefusatti10}. 
 
We have checked that this mismatch is not due to the fact that  the two simulations have different resolution but to the fact that different triangle configurations are sampled. When the triangle configurations and scales  coincide, we do not observe any mismatch. 
Note also that in  the cases where a mismatch appears --due to different configurations being sampled--,  it is also present  (and quantitatively similar) in the the 1-loop theory prediction (see for example top left panel of Fig. \ref{Qz1}).
 
At $z=0$, the deviation of our model from the data is typically less than 5\% and always less than 10\% for $k<0.4$ $h$/Mpc, whereas for SC the deviation reaches values of up to $20\%$  and  1-loop clearly breaks down for $k\gtrsim0.1\, h/\mbox{Mpc}$.
 We also observe that for triangles close to equilateral, both the SC approach and our work present maximum differences to the N-body data. This might have to do with the fact that for equilateral (or close to equilateral) triangles the 3 sides enter the non-linear regime at the same time, and thus,  non-linearities play a stronger role than for other triangle configurations, where each side enters the non-linear regime at different redshifts. Because of this, for other configurations the differences are smaller and remain within 5\% deviation.
 
 For $z=1$ all models work better because non-linearities do not play such an important role. However, even in this regime, our model works better than the other two. Again, for triangle configurations close to equilateral, our model reaches its maximum deviation of about 10\%. At $z=1$, all other triangle configurations typically have errors within 5\%.
 
As a cross-check, in Appendix \ref{Appendix_B} we compare our model with another set of simulations of non-standard LCDM model. In particular for equilateral configurations, our formula reaches deviations up to $10\%$. However for the scalene configurations $k_2=2k_1$ the deviations are only of order 3\%. In the cases studied here our model improves significantly  the SC fitting formula.

\section{Conclusions}
In this paper we propose a new simple formula to compute the dark matter bispectrum in the moderate non-linear regime ($k<0.4\, h/\mbox{Mpc}$) and for redshifts $z \leq1.5$. Our method is inspired by the approach presented in \cite{SC}, but includes a modification of the original formulae, namely Eq. \ref{abc_new}, and a prescription to better describe the BAO oscillations. The cosmology dependence of the reduced bispectrum is known to be very weak, and that of the bispectrum is almost completely contained in  the power spectrum. Given that the cosmological model today is well constrained by observations we have considered a single cosmology here.

Using LCDM simulations we fit the free parameters of our model. We end up with a simple analytic formula that is able to predict accurately the bispectrum for a LCDM Universe including the effects of BAO. 
Our main results are summarized by Eq. \ref{SC_formula} where the kernel is given by Eq. \ref{SC_kernel}, the functions $a,b,c$ are now given by $\tilde{a}, \tilde{b}, \tilde{c}$ of Eq. \ref{abc_new}, the fitting coefficients take the values reported in Table \ref{tabla_fit} and  the function $Q_3(n)$ is still given by Eq. \ref{eq:Q3}. The local slope of the linear power spectrum $n(k)$ is not  any more given directly by Eq. \ref{n} but is a smoothed (BAO-free, but with the same broadband behavior) function of $k$.

The main conclusions of our work are listed below.

\begin{enumerate}

\item Our method is able to predict the dark matter bispectrum for a wide range of triangle configurations up to $k=0.4\, h/\mbox{Mpc}$ and for a redshift range $0 \le z\le 1.5$. In particular, for the reduced bispectrum, our fitting formula agrees within 5\% with N-body data for most of the triangle configurations and always within 10\% for the worst cases. This presents a considerable improvement over previous phenomenological approaches and over the prediction of Eulerian perturbation theory.

\item The equilateral and quasi-equilateral configurations are the ones for which our model deviates most strongly from N-body data. We interpret this as being due to the fact that when the 3 sides of the triangle are similar, non-linearities start to play a role at the same time, and thus, the effect on the bispectrum is stronger than when the non-linearities enter at different times, i.e. for elongated triangles. Other methods, like the one described by SC show the same behavior.  

\item We have checked that our model also works well for non-standard LCDM cosmologies (see Appendix \ref{Appendix_B}). In particular, we have checked that for $k_2=2k_1$ the deviation between N-body data and our model is never higher than 3\% and for equilateral triangles reaches 10\%.  Also in these non-standard LCDM cases studied here, our model works better than the SC fitting formula.

\end{enumerate}

We envision that this new analytic fitting formula will be very useful in providing a reliable prediction for the non-linear dark matter bispectrum for LCDM models. In particular, simple analytic predictions with high accuracy will be needed for the data analysis in the forthcoming era of precision data.

\section{Acknowledgments}
We thank Fabian Schmidt for providing the non-standard LCDM simulations used in Appendix \ref{Appendix_B}. H\'ector Gil-Mar\'in thanks the Argerlander Institut f\"{u}r Astronomie at the University of Bonn for hospitality. H\'ector Gil-Mar\'in is supported by CSIC-JAE grant. Christian Wagner and Licia Verde acknowledge support of  FP7-IDEAS-Phys.LSS 240117.

\appendix
\section{Appendix: bispectrum estimator \& error bars}\label{Appendix_A}

Here we present details on the computation of the bispectrum and its error bars from N-body simulations. Moreover, we compare our error estimates with the Gaussian analytic predictions and discuss the differences. 

We start by defining the estimator for the bispectrum as, 
\begin{equation}
\hat{B}({\bf k}_1,{\bf k}_2,{\bf k}_3)\equiv\frac{V_f}{V_B}\int_{k_1}d^3q_1\int_{k_2}d^3q_2\int_{k_3}d^3\delta_D({\bf q}_1+{\bf q}_2+{\bf q}_3)\delta_{q_1}\delta_{q_2}\delta_{q_3}
\end{equation}
where $V_f=(2\pi)^3/L_b^3\equiv k_f^3$ is the volume of the fundamental cell, $k_f$. The integration is defined over the bin $k_i-\Delta k_i/2 <q_i<k_i+\Delta k_i/2$. In this paper we always take $\Delta k=3 k_f$. $V_B$ is the six-dimensional volume of triangles defined by the triangle sizes $k_1$, $k_2$ and $k_3$ with uncertainty $\Delta k$. Its value can be approximated by
\begin{equation}
V_B(k_1,k_2,k_3)=\int_{k_1} d^3q_1\int_{k_2} d^3q_2\int_{k_3} d^3 q_3\, \delta_D({\bf q}_1+{\bf q}_2+{\bf q}_3)\simeq 8\pi^2 k_1 k_2 k_3 \Delta k^3
\end{equation}
which is good enough for not too small values of $k_i$. The variance associated to this estimator depends on higher-order correlation functions: up to the 6-point connected correlation function. However, the main contribution to the variance is given by the power spectrum. Assuming that the fields are Gaussian, the variance associated to estimator presented above is \citep{error},
\begin{equation}
\Delta\hat B^2({\bf k}_1,{\bf k}_2,{\bf k}_3)=s_B \frac{V_f}{V_B}(2\pi)^3P(k_1)P(k_2)P(k_3)
\label{th_variance}
\end{equation}
where the symmetry factor is $s_B=6,\, 2,\, 1$ for equilateral, isosceles or scalene configurations. The factor $(2\pi)^3$ comes from our definition of the power spectrum and bispectrum in Eq. $\ref{Pk}$ and $\ref{Bk}$.

On the other hand, the discretized version of this estimator used in this paper is, $\widetilde{B}$
\begin{equation}
\widetilde B({\bf k}_1,{\bf k}_2,{\bf k}_3)=\frac{L_b^6}{N_{tri}}\sum_j^{N_{tri}}\mbox{Re}\left[\delta^d_j({k_1})\delta^d_j({k_2})\delta^d_j({k_3})\right]
\label{bispectrum_estimator}
\end{equation}
where $N_{tri}$ is the number of random triangle configurations used to compute the bispectrum; and $j$  runs over these triangle configurations. For this work we use a number of random triangles that increases with $k$ in the same way as the number of fundamental triangles: $\sim\tilde V_B/V_f^2$. It reaches up to $N_{tri}\sim10^9$ for scales $k\sim0.4\,h/$Mpc. We have checked that increasing the  number of random triangles beyond this value has no effect neither in the value of the bispectrum nor in its error. The index $d$ in the $\delta$ field stands for a discrete and dimensionless quantity. Therefore the quantity $\mbox{Re}\left[\delta^d_j({k_1})\delta^d_j({k_2})\delta^d_j({k_3})\right]$ needs to be rescaled with the factor $L_b^6$ to make $\widetilde B$ matching with the definition of the bispectrum in Eq. \ref{Bk}.
We compute the variance of this estimator $\widetilde{B}$ by the sample variance derived from the $N_r$ realizations,
\begin{equation}
{\Delta \widetilde B}^2({\bf k}_1,{\bf k}_2,{\bf k}_3)=\frac{1}{N_r-1}\sum_i^{N_r}\left(\widetilde B_i({\bf k}_1,{\bf k}_2,{\bf k}_3)-\langle\widetilde{B}({\bf k}_1,{\bf k}_2,{\bf k}_3) \rangle\right)^2
\label{sims_variance}
\end{equation}
where $\widetilde B_i$ is the bispectrum derived from the realization $i$ and $\langle\widetilde{B}\rangle$ is the mean over all realizations,
\begin{equation}
\langle\widetilde{B}({\bf k}_1,{\bf k}_2,{\bf k}_3)\rangle\equiv\frac{1}{N_r}\sum_i^{N_r}\widetilde B_i({\bf k}_1,{\bf k}_2,{\bf k}_3)\,.
\label{bisp_mean}
\end{equation}
The error on the mean $\langle \tilde{B} \rangle$ is then simply given by 
\begin{equation}
\sigma_{\langle \tilde{B} \rangle}={\Delta \widetilde{B}}/\sqrt{N_r}\,.
\label{error_mean}
\end{equation}

When comparing the measured N-body bispectrum with theoretical models and also when comparing Eq. \ref{th_variance} with Eq. \ref{sims_variance}, it is important to take into account the effect on the finite size of the triangle bins: each configuration is defined in terms of the sides of the triangle $k_i\pm \Delta k/2$. In this case, we are assuming $\Delta k=3 k_f$ and a large number of fundamental triangles fit into this bin. For certain configurations, it turns out that we have more triangles with $k$ larger than the central value, $k_i$. Because of that, one must correct  the sides of the triangles by
\begin{equation}
\tilde{k_i}=\frac{1}{N_{tri}}\sum_j^{N_{tri}}k_{i}^j
\label{k_correction}
\end{equation}
where $i=1,2,3$ for each dimension and the sum is taken over all random triangle generated in the bin.
This correction is extremely important at large scales and for very squeezed triangles, and less important for equilateral configuration.

\begin{figure}
\centering
\includegraphics[clip=false, trim= 110mm 45mm 23mm 45mm,scale=0.35]{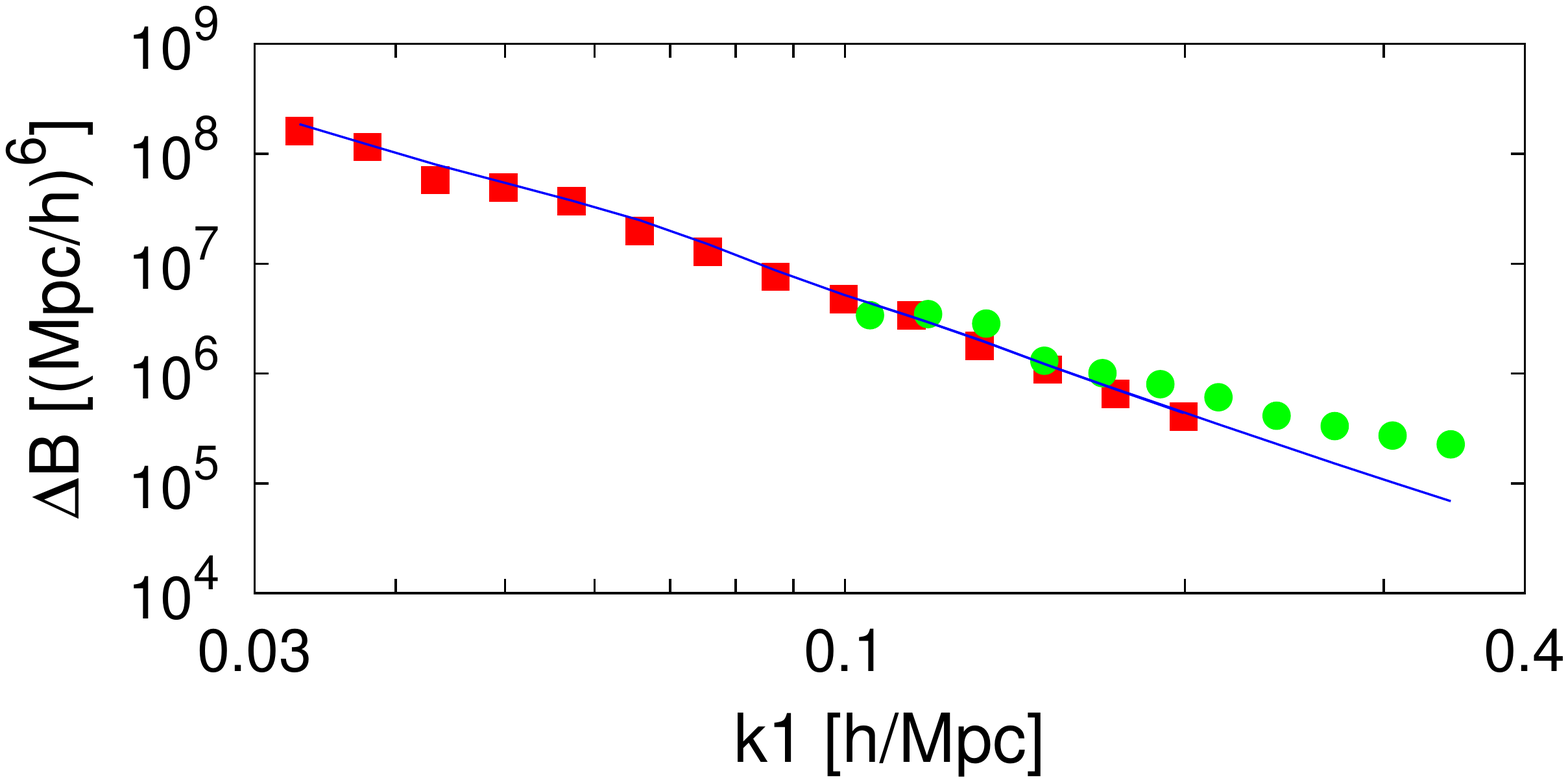}
\includegraphics[clip=false,trim= 23mm 45mm 110mm 45mm, scale=0.35]{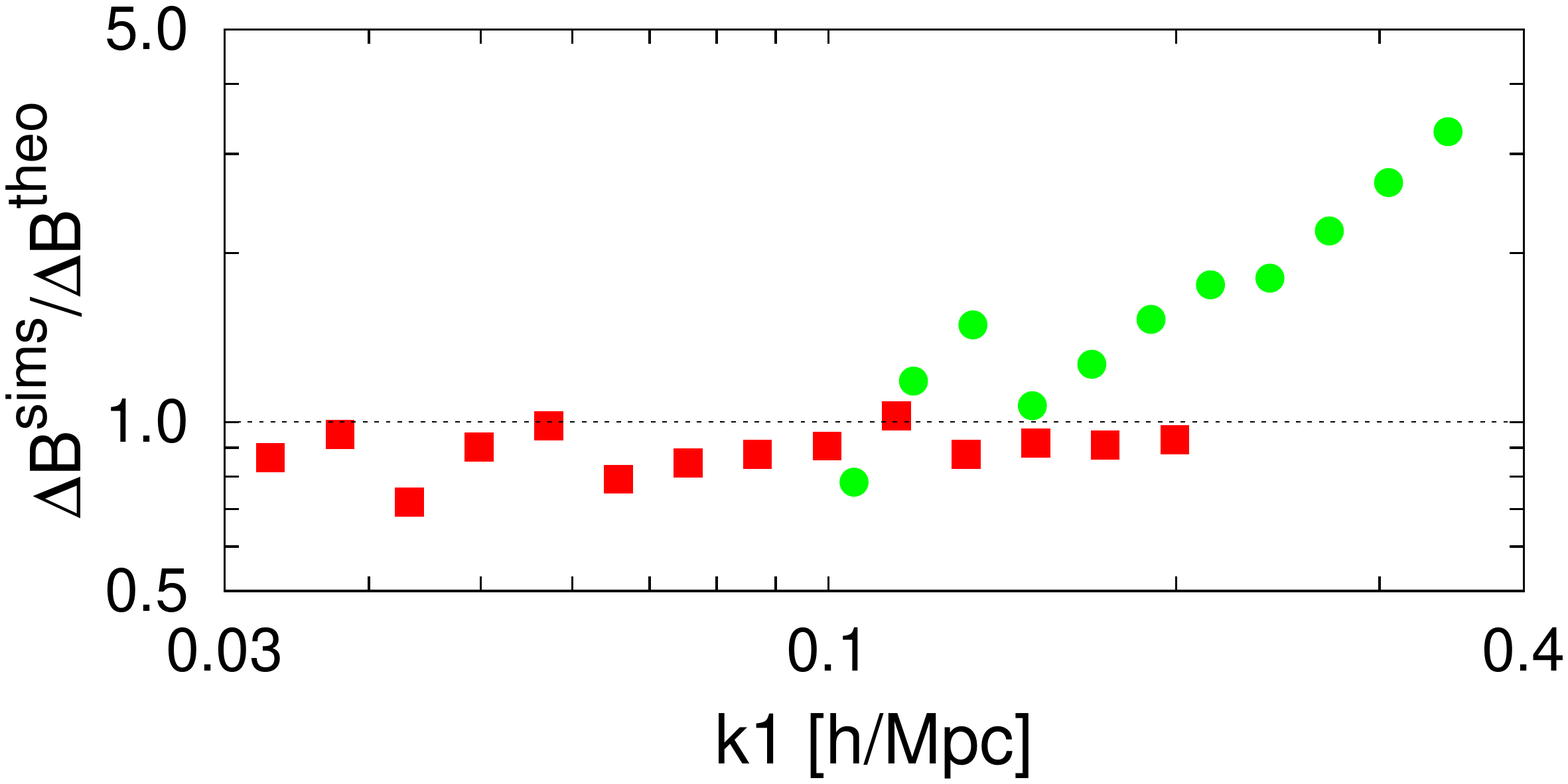}

\caption{{\it Left panel}: $\Delta B$ for $k_2/k_1=1$ and $\theta_{12}=0.6\pi$ triangles as a function of $k_1$ derived from simulations A (red squares) and from simulations B (green circles). Blue line is the theoretical prediction according to Eq. \ref{th_variance}. {\it  Right panel}:   ${\Delta \tilde B}/{ \Delta \hat B}$ for simulations A (red squares) and B (green circles) }
\label{plot_equilateral}
\end{figure}

In Fig \ref{plot_equilateral} we present a comparison of the error estimation from theoretical models (Eq. \ref{th_variance}) and simulations (Eq. \ref{sims_variance}). In the left panel we show the error of the bispectrum associated to a volume of 1 single box for $\Delta k=3k_f$: using the theoretical model and the simulations for the case of $k_2/k_1=1$ and $\theta_{12}=0.6\pi$ triangle configurations. The blue line shows the theoretical model prediction for the error of the bispectrum of one single realization using the non-linear power spectra from simulations A and B (Eq. \ref{th_variance}); whereas the red squares and green circles show the dispersion among the runs of simulations A and B respectively (Eq. \ref{sims_variance}). In the right panel the ratio between the errors according to the simulations and the Gaussian prediction is plotted for simulations A (red squares) and B (green circles). The error estimates of simulations A agrees well with the theoretical model. On the other hand, on small scales the error estimates of simulations B is larger than the theoretical model and further increase with decreasing scale. Similar results were found by \cite{guo1}. These differences are due to the fact that Eq. \ref{th_variance} neglects any higher-order contributions (because it assumes Gaussianity). However, at small scales this is no longer a good approximation as it is shown in \cite{sefusatti06}. Furthermore, the errors of simulations B have been estimated by dividing each of the 3 simulation boxes into 8 sub-boxes. This introduces extra non-Gaussian terms \citep{sefusatti06} that are not taken into account in Eq. \ref{th_variance}. 

\section{Appendix: our fitting formula for non-standard LCDM models}\label{Appendix_B} 

Here we test how our model works with different LCDM simulations to those we have used to fit the $a_i$ parameters. In particular, we test our model with LCDM simulations with a $f(R)$-like power spectrum. For a full description of the simulations and the $f(R)$ gravity we refer the reader to \cite{hgm}. These simulations were run with ENZO code and have slightly different cosmology than the ones used in the rest of this paper: $\Omega_\Lambda=0.76$, $\Omega_m=0.24$, $\Omega_b=0.04181$, $h=0.73$. They consist of 6 realizations that  contain $256^3$ particles in a box of $400$ Mpc/$h$ per side. The one-quarter Nyquist frequency is $k_{N}/4=0.5$ $h$/Mpc.

\begin{figure}
\centering
\includegraphics[clip=false, trim= 110mm 33mm 25mm 35mm,scale=0.35]{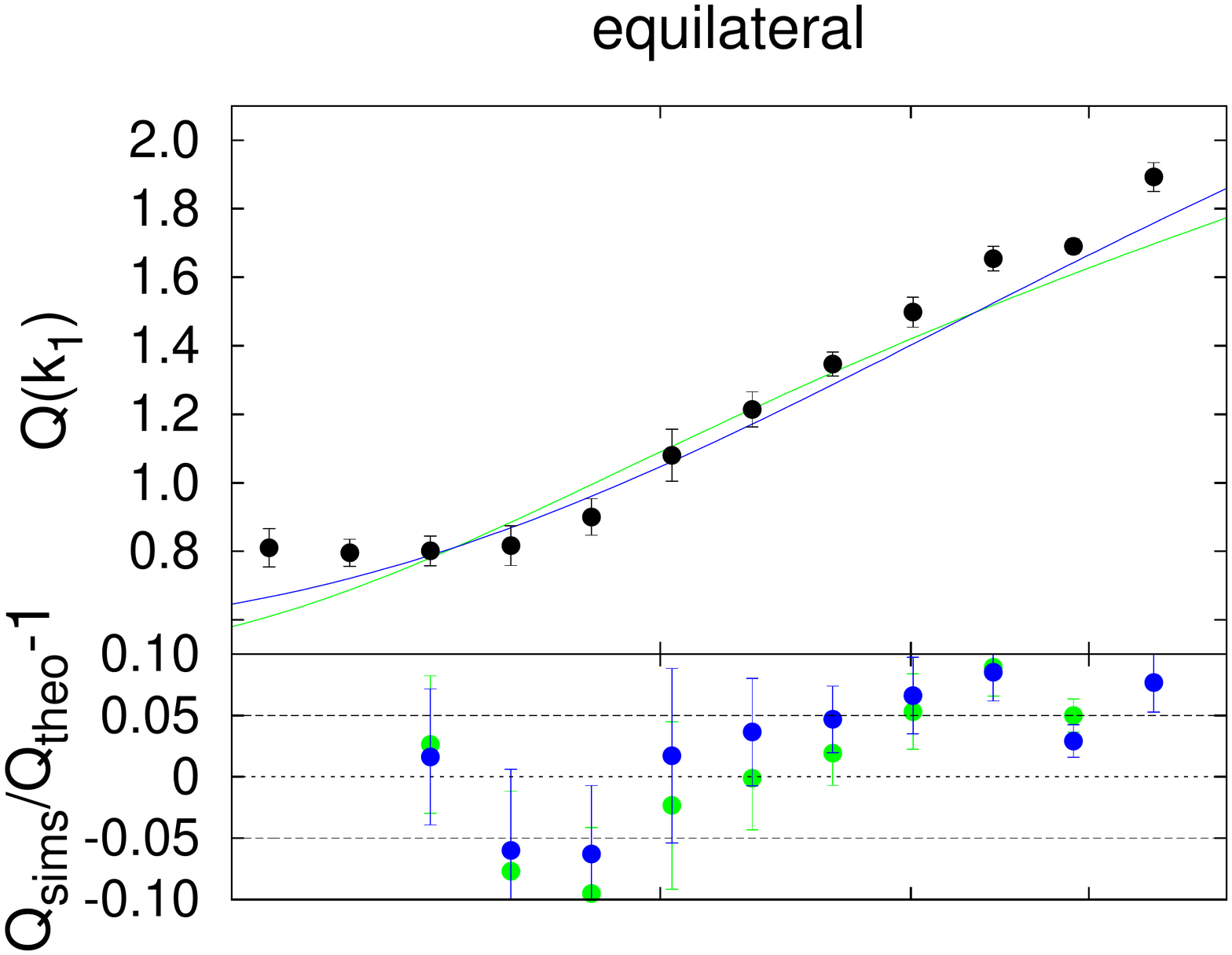}
\includegraphics[clip=false,trim= 25mm 33mm 110mm 33mm, scale=0.35]{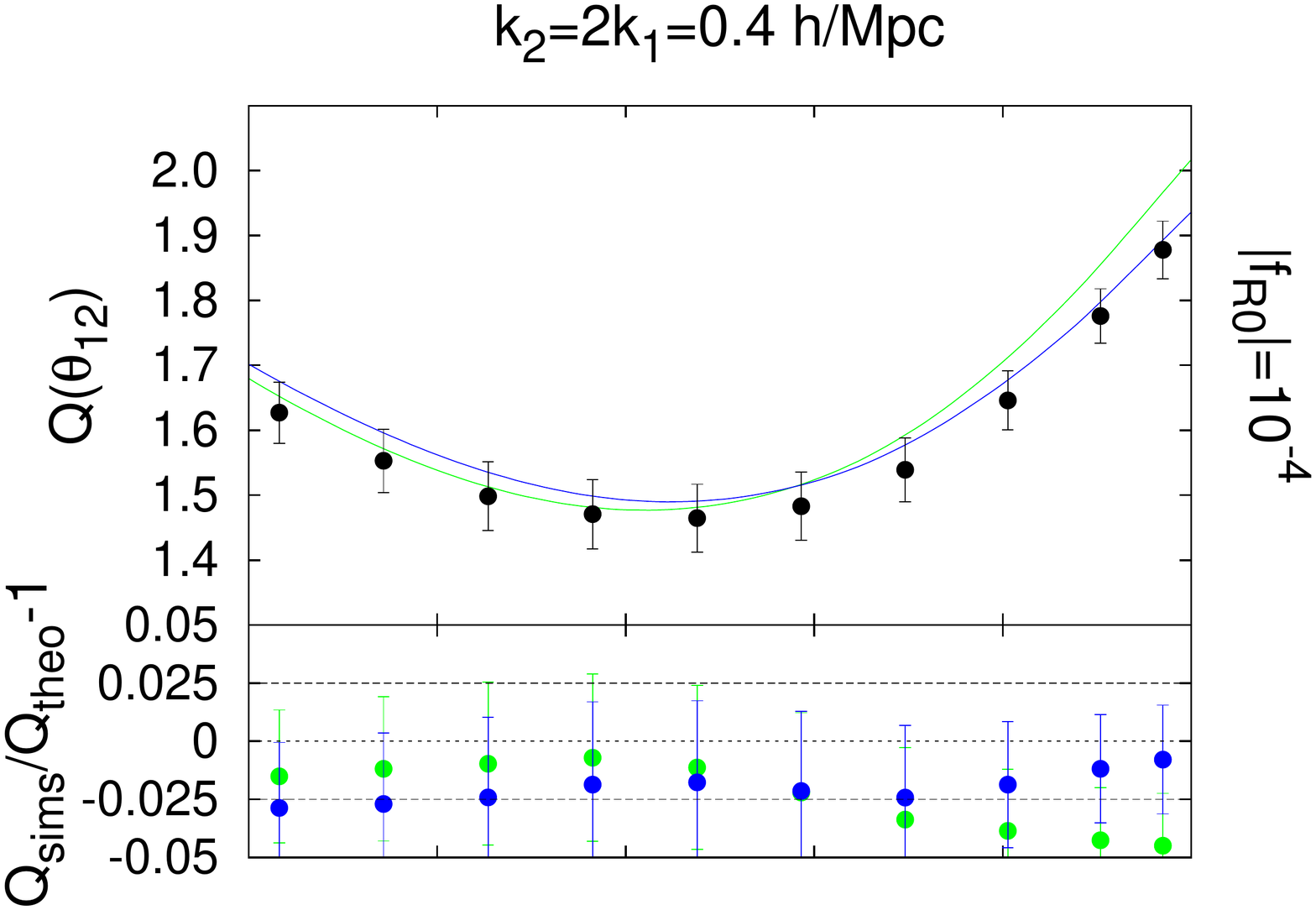}

\includegraphics[clip=false, trim= 110mm 33mm 25mm 33mm,scale=0.35]{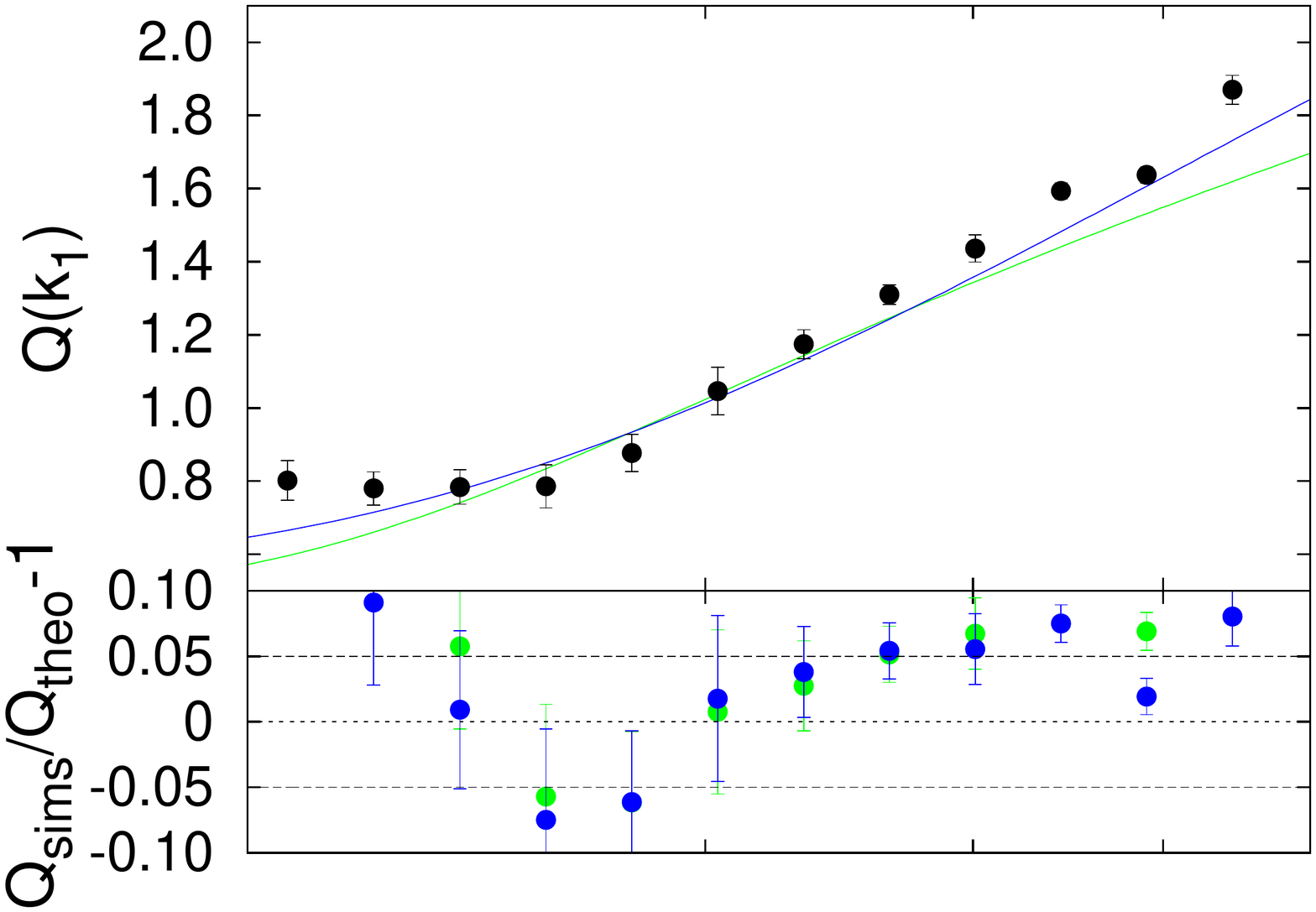}
\includegraphics[clip=false,trim= 25mm 33mm 110mm 33mm, scale=0.35]{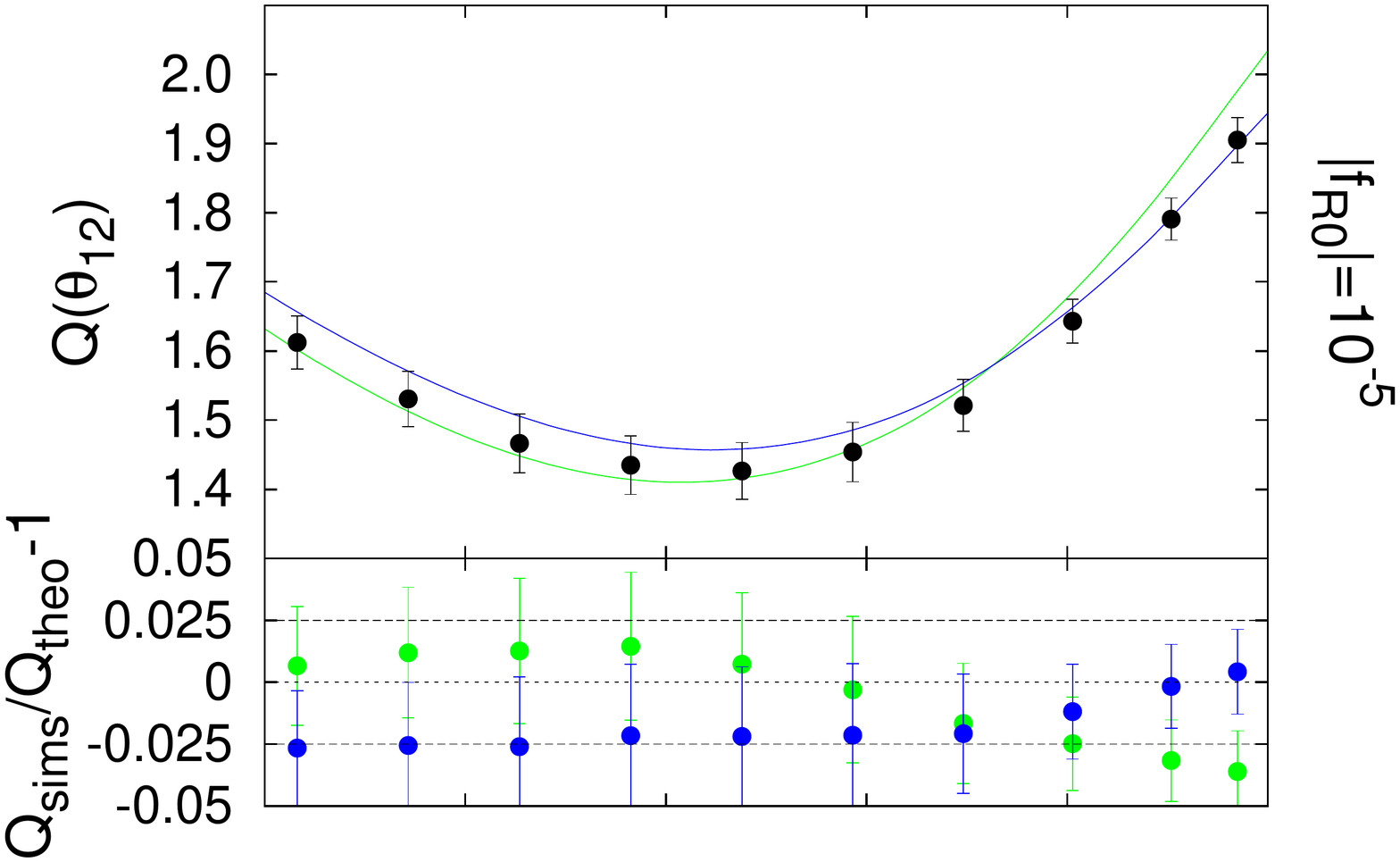}

\includegraphics[clip=false, trim= 110mm 33mm 25mm 33mm,scale=0.35]{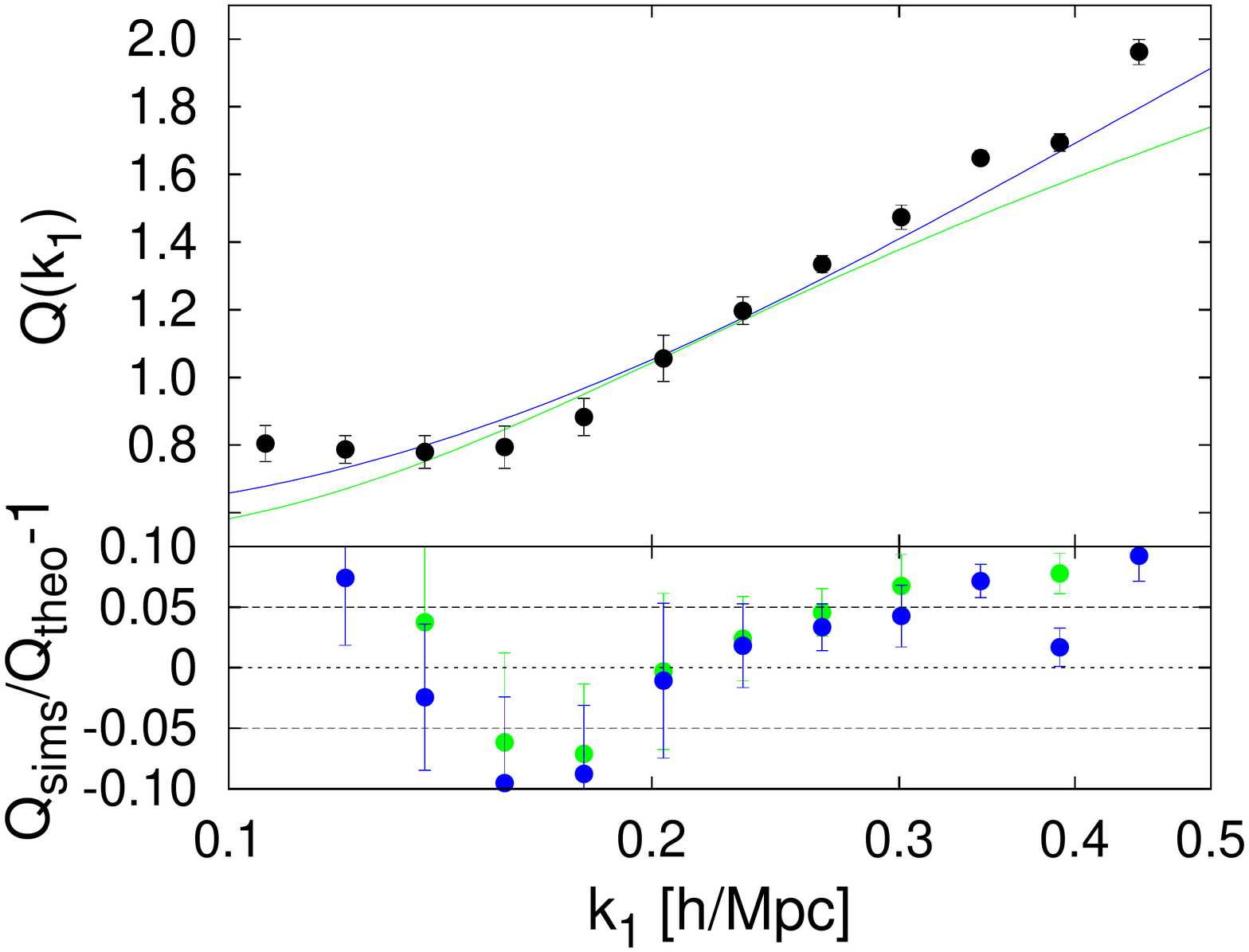}
\includegraphics[clip=false,trim= 25mm 33mm 110mm 33mm, scale=0.35]{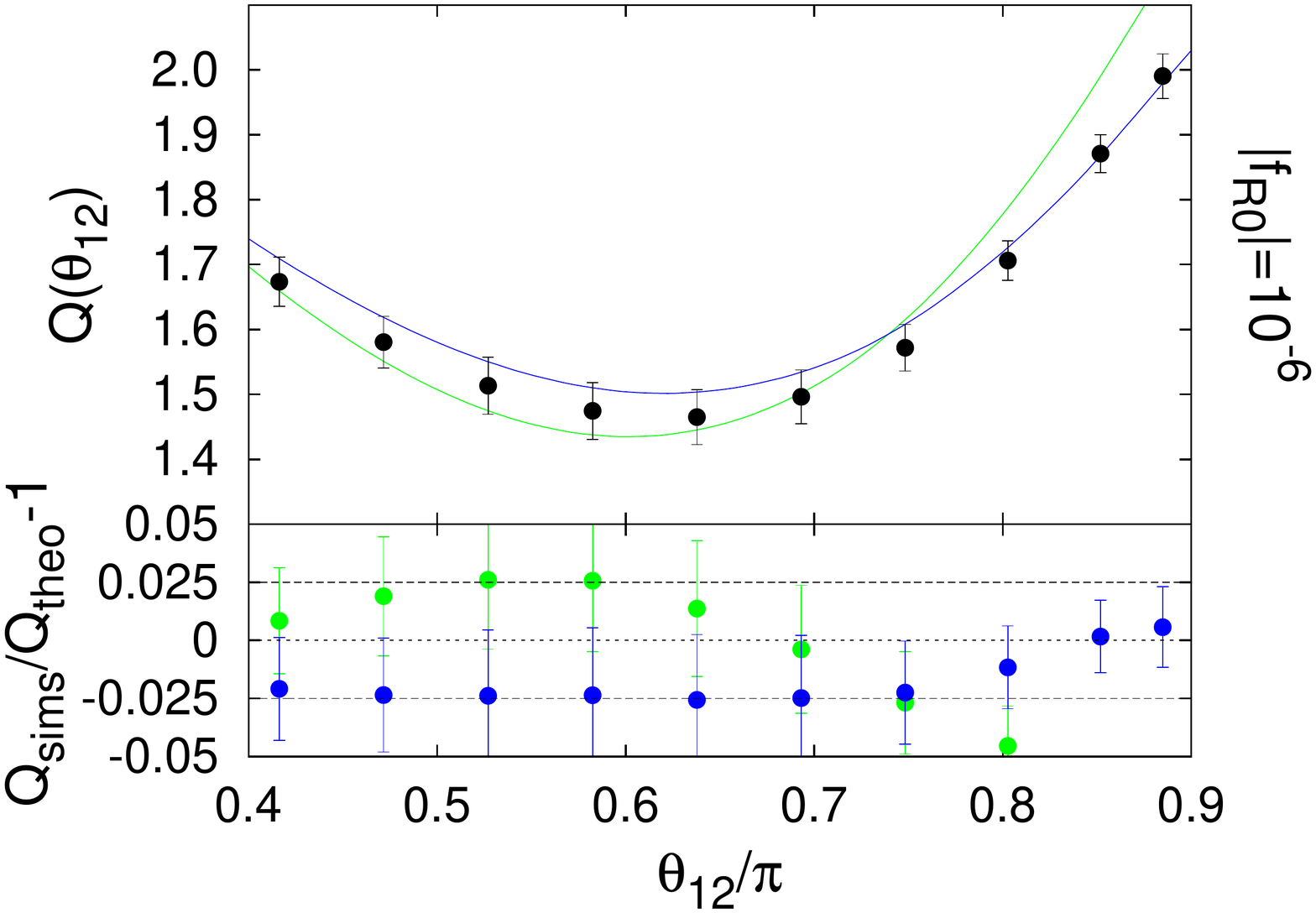}

\caption{In the left panels: $Q$ vs. $k_1$ for equilateral configuration. In the right panels  $Q$ vs $\theta_{12}/\pi$ for $k_2=2k_1=0.4\, h/\mbox{Mpc}$. From top to bottom we show different non-standard LCDM models. All the panels correspond to LCDM models with $f(R)$-like power spectrum from \cite{hgm}. Top panels correspond to matching to $|f_{R0}|=10^{-4}$; middle panels to $|f_{R0}|=10^{-5}$; bottom panels to $|f_{R0}|=10^{-6}$. The {\it sub-panel} shows the corresponding ratio between simulations and different theoretical models: SC (green points) and our work (blue points). In the right sub-panels, dashed lines mark 2.5\% deviation, whereas in the left panels mark the 5\% deviation.
The bispectrum and its error are estimated by Eq. \ref{bisp_mean} and \ref{error_mean} (see Appendix \ref{Appendix_A}).
}

\label{mod_GR}

\end{figure}

In Fig. \ref{mod_GR} the reduced bispectrum $Q$ is shown: in the right panel as a function of the angle between ${\bf k}_1$ and ${\bf k}_2$, namely $\theta_{12}$, for $k_2=2k_1$; in the left panel as a function of $k_1$ for equilateral configuration, both for $z=0$.
All panels correspond to a LCDM model with no BAOs whose initial conditions make their power spectrum look like a $f(R)$-like one. In order to do that a running index has been adopted in the initial conditions (see Table 1 in \cite{hgm} for details). Panels correspond to LCDM simulations that match with $f(R)$ models whose $|f_{R0}|$ parameter\footnote{see Eq. 2.3 in \cite{hgm} for a definition of $f_{R0}$} is: $10^{-4}$ (top panels), $10^{-5}$ (middle panels) and $10^{-6}$ (bottom panels ).
 
Black points are data from simulations, green line is the SC prediction and blue line the prediction of our model. In the right panel we only compare data points for  $0.4<\theta_{12}/\pi<0.9$  and in the left panel $0.1\, h/\mbox{Mpc}<k<0.5\, h/\mbox{Mpc}$ in order to ensure that all the $k_i$ are smaller than a quarter of the Nyquist frequency. 

Considering the right panels of Fig. \ref{mod_GR} ($k_2=2k_1=0.4\, h/\mbox{Mpc}$), we see that our model describes the data within about 3\%. In particular for small scales ($\theta_{12}<0.7\pi$), both SC and our model agree with the simulations data well; however at large scales ($\theta_{12}>0.7\pi$), our model fits the data points better. 
In the left panel of Fig. \ref{mod_GR} (equilateral configuration), we see that our model shows deviations up to 10\%. As for the standard LCDM model, the equilateral configuration is the one with the largest deviations. However even in this case, our formula behaves better than the SC model, especially at small scales. 

Therefore we conclude that our formula is general enough to be applied also to some non-standard LCDM models and in particular works better than SC at small scales.

\section{Appendix: one-loop correction terms for the power spectrum and bispectrum}\label{Appendix_C}

Here we present a short description of the equations used to compute the one-loop correction in Eulerian perturbation theory for the bispectrum shown in Fig. \ref{Qz0} and \ref{Qz1}. For a detailed description of PT see \cite{PT_review,error}.

Up to one-loop, the power spectrum can be expressed as,
\begin{equation}
P(k)=P^{(0)}(k)+P^{(1)}(k)+\dots
\end{equation}
where $P^{(0)}(k)=P_L(k)$ is the linear term and $P^{(1)}(k)=P_{13}(k)+P_{22}(k)$ is the one loop correction. For Gaussian initial conditions, the one-loop term consist of two terms\footnote{The $(2\pi)^3$ in the denominator comes from the definition of the power spectrum and bispectrum in Eq. \ref{Pk} and \ref{Bk} }, 
\begin{eqnarray}
P_{22}&=&\frac{2}{(2\pi)^3}\int d^3{\bf q}\, {F^s_2}^2({\bf q},{\bf k-q})P_L(q)P_L(|{\bf k-q}|)\\
P_{13}&=&\frac{6}{(2\pi)^3}P_L(k)\int d^3{\bf q} F^s_3({\bf k},{\bf q}, {\bf -q}) P_L(q)
\end{eqnarray}
$P_{22}$ accounts for the mode coupling between waves with wave-vectors $\bf{k} -  \bf{q}$ and $\bf q$, whereas $P_{13}$  can be interpreted as the one-loop  correction to the linear propagator.

Similarly to the power spectrum, the bispectrum up to one loop consists of two terms,
\begin{equation}
B({\bf k}_1, {\bf k}_2, {\bf k}_3)=B^{(0)}({\bf k}_1, {\bf k}_2, {\bf k}_3)+B^{(1)}({\bf k}_1, {\bf k}_2, {\bf k}_3)+\dots
\end{equation}
For Gaussian initial conditions, the first non-zero term is the second-order contribution $B^{(0)}$ which is {\it tree level} correction, whereas $B^{(1)}$ is the  one-loop correction. The tree level term can be expressed as,
\begin{equation}
B^{(0)}({\bf k}_1, {\bf k}_2, {\bf k}_3)=2F^s_2({\bf k_1},{\bf k_2})P_L(k_1)P_L(k_2)+\mbox{2 cyc. perm.}
\end{equation}
whereas the one loop consist of four terms (only for Gaussian initial conditions),
\begin{eqnarray}
B^{(1)}({\bf k}_1, {\bf k}_2, {\bf k}_3)&=&B^{I}_{222}({\bf k}_1, {\bf k}_2, {\bf k}_3)+B�_{123}({\bf k}_1, {\bf k}_2, {\bf k}_3)+\\
\nonumber &+&B^{II}_{123}({\bf k}_1, {\bf k}_2, {\bf k}_3)+B^{I}_{114}({\bf k}_1, {\bf k}_2, {\bf k}_3)
\end{eqnarray}
Each of these terms read as,
\begin{eqnarray}
B_{222}^I&=& \frac{8}{(2\pi)^3} \int d^3{\bf q}\, F_2^s({\bf -q}, {\bf q} +{\bf k}_1)F_2^s(-{\bf q}-{\bf k}_1,{\bf q}-{\bf k}_2)F_2^s({\bf k}_2-{\bf q},{\bf q})P_L(q)\times\\
\nonumber &\times& P_L(|{\bf k}_1+{\bf q}|)P_L(|{\bf k}_2-{\bf q}|)\\
B_{123}^I&=& \frac{6}{(2\pi)^3}P_L(k_1)\int d^3{\bf q}\, F_3^s({\bf k}_1, {\bf k}_2-{\bf q},{\bf q})F_2^s({\bf k}_2-{\bf q},{\bf q})P_L(|{\bf k}_2-{\bf q})P_L(q)+\mbox{5 perm.}\\
B_{123}^{II}&=&F_2^s({\bf k}_1, {\bf k}_2)\left[P_L(k_1)P_{13}(k_2)+P_L(k_2)P_{13}(k_1)\right]+\mbox{2 perm.}\\
B_{114}^I&=&\frac{12}{(2\pi)^3}P_L(k_1)P_L(k_2)\int d^3 {\bf q}\, F_4^s({\bf q},-{\bf q},-{\bf k}_1,-{\bf k}_2)P_L(q)+\mbox{2 perm.}
\end{eqnarray}
where $F^s_i$ are the {\it symmetrized} kernels. For an EdS Universe, the non-symmetric kernels read as, 

\begin{eqnarray}
F_n({\bf q}_1,\dots,{\bf q}_n)&=&\sum_{m=1}^{n-1} \frac{G_m({\bf q}_1,\dots,{\bf q}_m)}{(2n+3)(n-1)} \left[(2n+1)\alpha({\bf k},{\bf k}_1) F_{n-m}({\bf q}_{m+1},\dots,{\bf q}_n)+\right.\\
\nonumber &+&\left.2\beta({\bf k}, {\bf k}_1,{\bf k}_2)G_{n-m}({\bf q}_{m+1},\dots,{\bf q}_n)\right]\\
G_n({\bf q}_1,\dots,{\bf q}_n)&=&\sum_{m=1}^{n-1} \frac{G_m({\bf q}_1,\dots,{\bf q}_m)}{(2n+3)(n-1)} \left[3\alpha({\bf k},{\bf k}_1) F_{n-m}({\bf q}_{m+1},\dots,{\bf q}_n)+\right.\\
\nonumber &+&\left.2n\beta({\bf k}, {\bf k}_1,{\bf k}_2)G_{n-m}({\bf q}_{m+1},\dots,{\bf q}_n)\right]
\end{eqnarray}
with $F_1=G_1=1$. Also, ${\bf k}_1\equiv {\bf q}_1+\dots+{\bf q}_m$, ${\bf k}_2\equiv {\bf q}_{m+1}+\dots+{\bf q}_n$, ${\bf k}\equiv{\bf k}_1+{\bf k}_2$ and the functions $\alpha$ and $\beta$ are defined as,
\begin{eqnarray}
\alpha({\bf k}, {\bf k}_1)&\equiv&\frac{ {\bf k}\cdot{\bf k}_1}{k_1^2}\\
\beta({\bf k},{\bf k}_1,{\bf k}_2)&\equiv&\frac{k^2({\bf k}_1\cdot{\bf k}_2)}{2k_1^2k_2^2}
\end{eqnarray}
In order to obtain the symmetric kernels one has to symmetrize them with respect to their arguments,
\begin{equation}
F^s_n({\bf q}_1,\dots,{\bf q}_n)=\frac{1}{n!}\sum_\pi F_n({\bf q}_{\pi (1)},\dots,{\bf q}_{\pi (n)})
\end{equation}
where the sum is taken over all	 the permutations $\pi$ of	the set $\{1,\dots , n\}$.

Finally, the reduced bispectrum up to one loop can be written as,
\begin{equation}
Q=\frac{ B^{(0)}+B^{(1)}+\dots }{ \Sigma^{(0)}+\Sigma^{(1)}+\dots }\simeq Q^{(0)}+Q^{(1)}+\dots
\end{equation}
where
\begin{eqnarray}
\Sigma^{(0)}&=&P_L(k_1)P_L(k_2)+P_L(k_1)P_L(k_3)+P_L(k_2)P_L(k_3)\\
\Sigma^{(1)}&=&P_L(k_1)P^{(1)}(k_2)+P^{(1)}(k_1)P_L(k_2)+P_L(k_1)P^{(1)}(k_3)+\\
\nonumber &+& P^{(1)}(k_1)P_L(k_3)+P_L(k_2)P^{(1)}(k_3)+P^{(1)}(k_2)P_L(k_3)
\end{eqnarray}
Then it is easy to show that the linear and 1-loop correction terms for $Q$ reads as,
\begin{eqnarray}
Q^{(0)}&=&\frac{B^{(0)}}{\Sigma^{(0)}} \\
Q^{(1)}&=&\frac{B^{(1)}-Q^{(0)}\Sigma^{(1)}}{\Sigma^{(0)}}
\end{eqnarray}

\end{document}